\documentclass[prd,twocolumn,aps,superscriptaddress,nofootinbib,showpacs]{revtex4}
\usepackage{amsmath}
\usepackage{amssymb}
\usepackage{graphicx}
\usepackage{color}
\usepackage{bm}
\usepackage{times}
\usepackage{hyperref}
\hypersetup{
  colorlinks=true,
  citecolor=blue,
  linkcolor=blue,
  urlcolor=blue}

\usepackage{epsfig}
\usepackage{dcolumn}
\usepackage{float}

\usepackage{soul}

\usepackage{epsfig}
\usepackage{dcolumn}
\usepackage{morefloats}
\def\be{\begin{equation}}
\def\ee{\end{equation}}
\def\bea{\begin{eqnarray}}
\def\eea{\end{eqnarray}}
\def\gsim{\ \rlap{\raise 2pt\hbox{$>$}}{\lower 2pt \hbox{$\sim$}}\ }
\def\lsim{\ \rlap{\raise 2pt\hbox{$<$}}{\lower 2pt \hbox{$\sim$}}\ }
\def\dslash{\kern-4pt \not{\hbox{\kern-2pt $\partial$}}}
\def\pslash{\not{\hbox{\kern-2pt p}}}

\newcommand{\dcp}{\delta_{CP}}
\newcommand{\nova}{NO$\nu$A\ }

\begin{document}

\renewcommand{\arraystretch}{2}
\DeclareGraphicsExtensions{.eps,.ps}


\title{The physics of antineutrinos in DUNE and determination of  octant
and $\dcp$ }



\author{Newton Nath}
\email[Email Address: ]{newton@prl.res.in}
\affiliation{
Physical Research Laboratory, Navrangpura,
Ahmedabad 380 009, India}
\affiliation{Indian Institute of Technology, Gandhinagar, Ahmedabad--382424, India}

\author{Monojit Ghosh}
\email[Email Address: ]{monojit@prl.res.in}
\affiliation{
Physical Research Laboratory, Navrangpura,
Ahmedabad 380 009, India}
  
\author{Srubabati Goswami}
\email[Email Address: ]{sruba@prl.res.in}
\affiliation{
Physical Research Laboratory, Navrangpura,
Ahmedabad 380 009, India}

\begin{abstract}

The octant of the leptonic mixing angle $\theta_{23}$ and the 
CP phase $\delta_{CP}$ are the two major unknowns (apart from neutrino mass hierarchy) in neutrino oscillation physics.  
It is well known that the precise determination of octant and $\delta_{CP}$ is interlinked 
through the octant-$\dcp$ degeneracy. In this paper we study the proficiency of the DUNE experiment 
to  determine these parameters 
scrutinizing, in particular, the role played by the 
antineutrinos, the broadband nature of the beam and the  matter 
effect. 
It is well known that  for $P_{\mu e}$ and $P_{\bar{\mu} \bar{e}}$ 
the octant-$\delta_{CP}$ degeneracy  occurs at different values of $\delta_{CP}$, combination of 
neutrino and antineutrino runs help to resolve this. However, in regions where neutrinos do not have octant degeneracy  
adding antineutrino data is expected to  decrease the sensitivity because of the degeneracy and reduced statistics. 
However we find that in case of DUNE  baseline, the antineutrino runs help even in parameter space where the antineutrino probabilities suffer from degeneracies. We explore this point in detail and point out that 
this happens because of the 
(i) broad-band nature of the beam  so that  even if there is degeneracy at
a particular energy bin, over the whole spectrum the degeneracy may not be 
there;  
(ii) 
the enhanced matter effect due to the comparatively longer baseline which creates an increased tension between the 
neutrino and the antineutrino probabilities which raises the  overall $\chi^2$  in case of combined runs. 
This feature is more prominent for IH  since the antineutrino probabilities in this case are much higher than 
the neutrino probabilities due to matter effects. The main role of antineutrinos in enhancing CP sensitivity is their ability to  
remove the octant-$\delta_{CP}$ degeneracy. However even if one assumes octant to be known the addition of antineutrinos can give 
enhanced CP sensitivity in some parameter regions due to the  tension between the neutrino and antineutrino 
$\chi^2$s.

\end{abstract}
\maketitle

\section{Introduction}

The discovery of a non-zero value of the 1-3 leptonic 
mixing angle $\theta_{13}$ by the reactor experiments 
have established the paradigm of oscillations of the neutrinos amongst 
three flavours on a firm footing.  The parameters involved are:  
two mass squared differences -- $\Delta m^2_{21}$, $\Delta m^2_{31}$, 
three mixing angles $\theta_{12}$, $\theta_{23}$ and $\theta_{13}$ 
and the CP violating phase $\delta_{CP}$.  
Among these $\Delta m^2_{21}$ and $\theta_{12}$ are measured  
by the solar neutrino and the KamLAND reactor neutrino experiments \cite{kamland}. 
The information on $\Delta m^2_{31}$ and $\theta_{23}$ has come from 
Super-Kamiokande (SK) \cite{sk_2010} atmospheric neutrino data, 
as well as  from the data 
of the  beam based experiments MINOS \cite{minos_latest} 
and T2K \cite{Abe:2014tzr}. 
The best-fit values and $3\sigma$ ranges of these parameters
are given in  \cite{Gonzalez-Garcia:2015qrr, Capozzi:2016rtj}
by  analyzing the global neutrino data. 
The remaining unknown oscillation parameters  are
(i) the sign of $|\Delta m^2_{31}|$ or the 
neutrino mass ordering. If we assume the neutrinos to be hierarchical then 
there can be two types of ordering -- the normal hierarchy (NH) corresponding 
to $m_1 \ll m_2 \ll m_3$ and $\Delta m^2_{31} > 0$  and the  
inverted hierarchy (IH) corresponding to $m_2 \approx m_1 \gg m_3$ and 
$\Delta m^2_{31} <0$,  (ii) the octant of $\theta_{23}$ -- with $\theta_{23} < 45^\circ$ corresponding to lower octant (LO) and $\theta_{23} > 45^\circ$ corresponding 
to higher octant (HO) and  (iii) the CP violating phase $\dcp$ for which the 
full range from $-180^\circ< \dcp< 180^\circ$ is still allowed 
at $3 \sigma$ 
C.L. \cite{Gonzalez-Garcia:2015qrr, Capozzi:2016rtj}. 
Information on these parameters can come from
the currently running superbeam experiments T2K \cite{t2k_dcphierocthint}
and \nova \cite{Adamson:2016tbq,Adamson:2016xxw}. 
However this is possible only for favourable values of parameters. 
The main problem which these experiments can face is due to 
parameter degeneracies by which it is meant that same parameters 
giving  equally good fit to the data. 
With $\theta_{13}$ unknown, an eight-fold degeneracy was identified 
which would make the precise determination of parameters difficult 
\cite{barger}. 
These were intrinsic $\theta_{13}$ degeneracy \cite{degeneracy3}, hierarchy-$\dcp$ degeneracy \cite{degeneracy2}
and octant degeneracy \cite{lisidegen}. 
With the precise determination of $\theta_{13}$ \cite{t2k_t13_jun2011, dchooz_latest,An:2015rpe,reno_t13} and inclusion of 
spectral information  the intrinsic degeneracy is now solved. 
However the lack of knowledge of hierarchy, octant and $\dcp$ 
can still give rise to degenerate solutions which can affect the 
sensitivities of these experiments towards these parameters 
\cite{novat2k,suprabhoctant,minakata_cp,Coloma:2014kca,Ghosh:2015ena}. 

In this paper our focus is on the determination of the octant of 
$\theta_{23}$ and the CP phase $\dcp$. 
Currently the most precise  measurements of the  parameter 
$\theta_{23}$ comes from the T2K experiment. 
The primary channel for this is the survival probability 
$P_{\mu \mu}$. 
For baselines shorter than $1000$ km 
this probability is a  function of $\sin^2 2\theta_{23}$ to the leading order
and suffers from an intrinsic octant degeneracy 
which refers 
to the same value of probability for $\theta_{23}$ and $\pi/2 - \theta_{23}$. 
The leading order term of the 
appearance channel probability $P_{\mu e}$ depends on the combination
$\sin^2 \theta_{23} \sin^2 2\theta_{13}$.  Although this does not exhibit 
intrinsic octant degeneracy, there can be uncertainties due to the 
$\sin^2\theta_{13}$ factor. It was shown  in \cite{hubercpv, cpcombo_sugiyama} that combining 
the reactor measurement of $\theta_{13}$ with the accelerator data will 
be helpful for extraction of information on octant from this channel. 
Thus, the precise measurement of $\theta_{13}$ from the reactor experiments
is expected to enhance the octant sensitivity coming from this channel. 
The combination of the disappearance and appearance channel measurements in 
long baseline experiments can also be helpful in resolving octant degeneracy 
because of the different functional dependence of the two probabilities 
on $\theta_{23}$. This creates a synergistic effect so that the octant 
sensitivity of both channels combined is higher 
\cite{octant_atmos,suprabhoctant,Coloma:2014kca}.  
T2K collaboration has performed 
a full three flavour analysis using information
from both ($\nu_\mu - \nu_\mu$) and ($\nu_\mu - \nu_e$) channels. 
They obtain best-fit $\sin^2 \theta_{23} \sim 0.52$ with a 
preference for NH \cite{t2k_dcphierocthint}.  
MINOS collaboration has also completed their combined analysis of 
disappearance and appearance data and have also included atmospheric neutrino 
data in their analysis \cite{minos_latest}. They get a best-fit at 
$\sin^2 \theta_{23} = 0.41$ for  IH. The first \nova disappearance results 
with $2.74 \times 10^{20}$ protons on target, give best-fit of $\sin^2\theta_{23} = 0.43 \oplus 0.60$
\cite{Adamson:2016xxw}.
The  latest analysis of Super Kamiokande atmospheric  neutrino 
data shows a weak  preference for NH-HO \cite{sk3gen}. 
Global analysis of neutrino data including all the different information 
gives the best-fit in LO for NH and in HO for IH
\cite{Gonzalez-Garcia:2015qrr, Capozzi:2016rtj}. 
Thus it is clear from the above discussion that at 
present the situation regarding octant of $\theta_{23}$ is quite intriguing. 

There have been studies on the  possibility of determining the octant 
from combined study of the experiments T2K and \nova using
their full projected exposure \cite{suprabhoctant, octant_atmos}. 
It was observed that the main problem in octant resolution 
arises due to the unknown value of $\dcp$ in the subleading terms of 
$P_{\mu e}$ which gives rise to octant-$\dcp$ degeneracy. 
Also, the lack of knowledge about hierarchy can create further problem 
with the occurrence of wrong hierarchy - wrong octant solutions 
\cite{Ghosh:2015ena}. 
Recently it was pointed out in \cite{suprabhoctant,Coloma:2014kca} 
that equal neutrino 
and antineutrino runs can help in resolving octant-$\dcp$ degeneracy. 
The reason being the octant-$\dcp$ combination suffering from degeneracy 
in neutrino probabilities are not degenerate for the antineutrino probabilities.
It was shown for instance in \cite{suprabhoctant} that combining T2K and \nova
running in equal neutrino and antineutrino mode for 2.5 years each and 3 years
each respectively can identify the correct octant at $2\sigma$ C.L.
irrespective of hierarchy and $\dcp$ if $\theta_{23} \leq 41^\circ$ 
or $\geq 49.5^\circ$. 

The degeneracies can also be alleviated if neutrinos pass through large
distances
in matter so that resonant matter effects develop. 
This is the case of the atmospheric neutrinos passing through matter. 
In this case the leading order term in $P_{\mu e}$ goes as  
$\sin^2 \theta_{23} \sin^2 2\theta_{13}^m$.
However, since at resonance $\sin^2 2\theta_{13}^m \approx 1$, the octant
degeneracy is resolved. Further, the  $P_{\mu \mu}$ channel also contains 
an octant sensitive term $\sin^4 \theta_{23} \sin^2 2\theta_{13}^m$ which
enhances the 
sensitivity  \cite{Choubey:2003yp}.
Octant sensitivity can also come from the $\Delta m^2_{21}$
dependent term which  
gives rise to an excess of  sub-GeV electron like events for 
the atmospheric neutrinos  
\cite{atmosoctant,GonzalezGarcia:2004cu}. 
In addition the antineutrino component in atmospheric neutrino flux can
also help in resolving octant ambiguity. 
It was shown that combined analysis of T2K and \nova with atmospheric neutrino 
data can give enhanced octant sensitivity \cite{octant_atmos}. 
The effect was found to be larger in multi-megaton water detectors 
like PINGU \cite{Choubey:2013xqa} or a LArTPC detector, sensitive to 
both muon and electron events \cite{octant_atmos}.

The current best-fit value for $\dcp$ is close to $-\pi/2$ although at 
$3\sigma$ C.L. the whole range of $[0,2\pi]$ remains allowed 
\cite{Capozzi:2016rtj, Gonzalez-Garcia:2015qrr}.
The $\dcp$ sensitivity of an experiment is often understood in terms of 
the CP asymmetry between the neutrinos and antineutrinos. 
\begin{equation} 
A_{cp} = \frac{P_{\mu e} - P_{\overline{\mu} \overline{e}}}{P_{\mu e} + P_{\overline{\mu} \overline{e}}}
\sim \frac{\sin\delta_{CP} }{\sin\theta_{13}}
\end{equation}  
However the diagnostics used for probing CP violation is the sum total
of the $\chi^2$ contribution of the neutrinos and antineutrinos:  
$\chi^2_{total} = \chi^2_{\nu} + \chi^2_{\bar{\nu}}$ which does 
not show the above dependence \cite{ourlongcp}. Hence 
one needs to understand the actual role played by antineutrinos, if any, 
for determination of CP violation. 
Indeed one already has a hint for  non-zero
$\dcp$ from only neutrino runs of T2K 
and \nova. 
Whereas the confirmation of CP violation independently from 
antineutrino runs in these experiments cannot be undermined, it has already 
been observed in the case of T2K that unless the parameter space contains 
octant degeneracy the antineutrinos do not play any role for discovery of CP
violation \cite{Ghosh:2015tan, jara_15}.  
However for the \nova experiments 
antineutrinos seem to be playing some role even when 
there is no octant degeneracy \cite{Ghosh:2015tan}. 
In this work, it is one of our goals to understand the role of     
antineutrinos for enhancing CP sensitivity for the DUNE baseline. 
In particular we explore whether the antineutrino runs can play any 
non-trivial contribution to the total $\chi^2$ if octant and hierarchy are 
assumed to be known and if so then what are the physics issues involved.

The current generation superbeam experiments T2K and \nova are off-axis
experiments using narrow band beams to reduce the backgrounds at the high 
energy tail. 
However the  future generation high statistics accelerator experiments   
plan to use on-axis configurations and high intensity wide band beams 
enabling them to explore oscillations over  a larger energy range. 
The examples for this are the European initiative LBNO 
and LBNE which was proposed in US using the FermiLab beamline. 
In 2014 it was proposed to combine these activities in a coherent 
international long-baseline neutrino program hosted at Fermilab with 
the detector at the Sanford Underground Research Facility (SURF) in South Dakota. 
On Jan. 30, 2015 the LBNE collaboration was officially dissolved, 
the new collaboration selected the name Deep Underground Neutrino Experiment
(DUNE).
 The baseline is 1300 km and the proposed detector is a 40 kt (or 34 kt)
modular Liquid Argon Time Projection Chamber (LArTPC) with the first 
phase being a 10 kt detector. There are several studies of the physics
prospects of a 1300 km baseline LArTPC using a wide band beam
\cite{lbne}.    
In particular octant and/or CP sensitivity of such a set-up has been 
considered 
in 
\cite{suprabhlbnelbno,sushant-lbne,Bora:2014zwa,raj_lbne1,raj_lbne2,Deepthi:2014iya}.   
In \cite{suprabhlbnelbno} the octant and CP sensitivity reach
of a 10 kt LArTPC detector for LBNE combined with T2K and \nova  was studied. 
In \cite{sushant-lbne} the minimum 
exposure for DUNE in conjunction with T2K, \nova and ICAL@INO experiment
was computed for giving a octant sensitivity with $\Delta \chi^2 = 25$ 
and a CP sensitivity with $\Delta \chi^2=9$ for 40\%  and $\sim$ 70\% 
coverage of $\dcp$.  
In \cite{Bora:2014zwa} the octant sensitivity of a 10 kt and 35 kt detector 
was studied with and without the near detector and also the role of 
precise knowledge of $\theta_{13}$ coming from reactor
experiments, in improving the sensitivities 
were studied.  
In \cite{raj_lbne1} octant and CP  sensitivity results were presented for 
a 10 kt detector and effect of including a near detector as well as
the role of atmospheric neutrinos were considered.  
In \cite{raj_lbne2} octant and CP sensitivity of a 35 kt detector, 
with and without magnetization, was studied. 
All these papers considered equal neutrino and antineutrino 
run for the octant sensitivity. 
Variation of proportion of neutrino and antineutrino run 
was studied in \cite{sushant-lbne} for only IH and $\theta_{23}=39^\circ$
for octant sensitivity and NH and $\theta_{23} = 51^\circ$ for 
CP sensitivity. This issue was also discussed in \cite{incremental} 
for a setup with a baseline of 1540 km 
where they concluded that for true hierarchy as 
NH equal neutrino and antineutrino run 
is better whereas for  true hierarchy as IH 30\% antineutrino run is optimal. 
These conclusions were drawn for true $\theta_{23} = 45^\circ$ and the 
results were presented in terms of fraction of $\dcp$ for which a $3\sigma$ 
signal of CP violation can be obtained.  

In this work,  our main goal to understand the role of     
antineutrinos for enhancing  octant and CP sensitivity for the DUNE baseline.
In particular, we study the impact of the broadband nature of 
the beam and the role of enhanced matter effects as compared to 
the currently running beam-based experiments T2K and \nova which have 
shorter baselines and hence less matter effects. 
To the best of our knowledge these features have not been emphasized 
earlier in the literature. In particular, a deeper understanding of the 
role played by antineutrinos will help  
in optimizing the amount of 
antineutrino run.    
We present the results of the octant sensitivity using different combinations 
of neutrino-antineutrino run 
(i) as a function of true $\dcp$ for fixed values of 
true $\theta_{23}$
(ii) 
as a function of true $\theta_{23}$ for fixed values of true $\dcp$ and 
(iii) also in the true ($\theta_{23}$ - $\dcp$) plane. 
These three kinds of plots allow us to study the dependendence of 
octant-sensitivity on these two parameters in an exhaustive 
manner. 
In addition
we present the allowed regions in the true-$\theta_{23}$ - test $\theta_{23}$ plane for both hierarchies and for true $\dcp = \pm 90^\circ$. 
These plots give the precision of $\theta_{23}$ at these values of $\dcp$. 
It is worthwhile to mention here that one of the main aims of the 
DUNE collaboration is to measure the parameter $\dcp$ which 
underscores the importance of combining neutrino and antineutrino runs. 
However the main role that the  antineutrinos play in
the determination of $\dcp$ is the 
removal of octant degeneracy and thus both issues are intimately connected.
To emphasize this point we also present the figures showing the CP discovery 
potential of DUNE for the cases when octant is assumed to be known and unknown. 
In particular we explore whether the antineutrino runs can play any 
non-trivial contribution to the total $\chi^2$ if octant and hierarchy are 
assumed to be known and if so then what are the physics issues involved.  
We  study how much  
fraction of antineutrino run  is optimum for values of $\theta_{23}$ in 
lower and  upper octants and in addition to the CP fractions, 
also show explicitly which are the 
CP values for which antineutrino run can be important. 
Note that most of the earlier works in literature have considered 
equal neutrino and antineutrino run for determination of 
octant and $\dcp$ in DUNE. 
We present the 
the results by varying the antineutrino component in the run. 
This constitutes another new feature of our study. 


The plan of the paper goes as follows: 
in the next section we give the experimental and simulation details of DUNE 
that have been taken into consideration. 
In Section (\ref{sec3}) we discuss the physics of the 
octant and CP sensitivity of the DUNE experiment in detail. 
 In Section (\ref{sec4}) we present our results. 
Section (\ref{sec4a}) contains the results for octant sensitivity and Section (\ref{sec4b}) is devoted for the discussions on CP sensitivity and role of antineutrinos in DUNE.
Finally we summarize and conclude in  Section (\ref{sec5}).
   
\section{ Experimental and Simulation Details }

In this paper  we have simulated the DUNE experiment
using the GLoBES package \cite{globes1,globes2}. 
In our simulation, we have considered a 10 kt configuration of the detector.  
This experiment is based on the existing Neutrinos at the Main Injector 
(NuMI) beamline design and the beam flux peaks at 2.5 GeV. 
Far detector will be located 4,850 feet underground.  
One of the options for DUNE is  
to have an initial  beam power of 
1.2 MW  which will be increased to 2.3 MW  later \cite{Acciarri:2015uup}.
In our simulation we consider neutrino flux \cite{cherdack} 
corresponding to 1.2 MW beam power which gives  $ 10^{21} $ protons on target (POT) per year. This corresponds to a proton energy of 120 GeV. 
In Table \ref{param_values} we list the representative values for the neutrino oscillation  parameters that we  have used in our numerical simulation.  
These values are consistent with the results  obtained from global-fit  
of world neutrino data. 
\cite{Gonzalez-Garcia:2014bfa,global_fogli,global_valle}.
Systematic errors are  taken into account using the method of
pulls \cite{pulls_gg, pull_lisi} as outlined in \cite{ushier}.
We have also added $5\%$ prior on  sin$^{2}2\theta_{13} $ in our numerical
simulation. The systematic errors and efficiencies corresponding to signal and background are taken from \cite{lbne}. Note that the values of these 
quantities given in \cite{Acciarri:2015uup} are somewhat different. Using these 
may change our numerical results to some extent though the 
main physics issues addressed in this work will not be altered.   

 \begin{table*}
 \begin{tabular}{|c|c|c|c|c|c|c|}
 \hline   
    & $ \sin^{2}2 \theta_{13} $ & $ \sin^{2}\theta_{12} $ & $ \theta_{23} $ & $\Delta m^2_{21}(eV^{2}) $ & $\Delta m^2_{31}(eV^{2}) $ & $ \delta_{CP} $ \\
 \hline
 \hline
  True Values & 0.1 & 0.31 & $35^\circ-55^\circ$ & 7.60 $ \times 10^{-5} $ &  2.40 $ \times 10^{-3}  $ & $ -180^\circ $ to $ +180^\circ $ \\
  \hline
  Test Values & 0.085 -- 0.115 & Fixed &$35^\circ-55^\circ$ & Fixed & (2.15 -- 2.65) $ \times 10^{-3} $ & $ -180^\circ $ to $ +180^\circ $ \\   
  \hline
 \end{tabular}
 \caption{Representative values of neutrino oscillation parameters.} 
 \label{param_values}
 \end{table*}

\section{Physics of octant sensitivity for a 1300 km baseline }
\label{sec3}
\begin{figure*}
        \begin{tabular}{lr}
                \hspace*{-0.35in} 
                \includegraphics[width=0.5\textwidth]{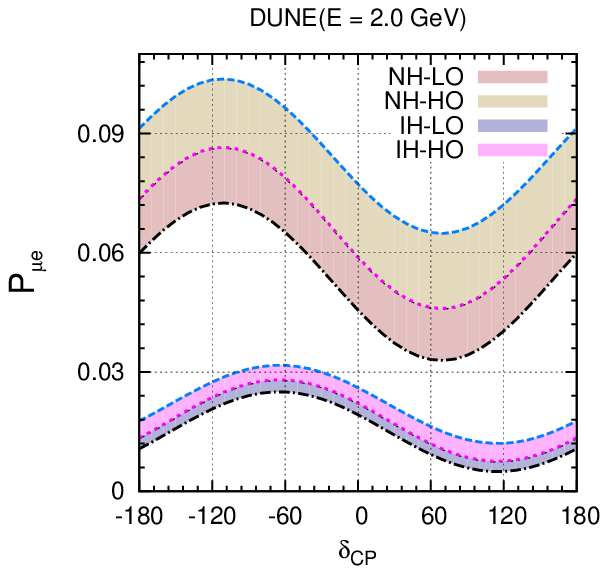}
                \hspace*{-1.0in}
                \includegraphics[width=0.5\textwidth]{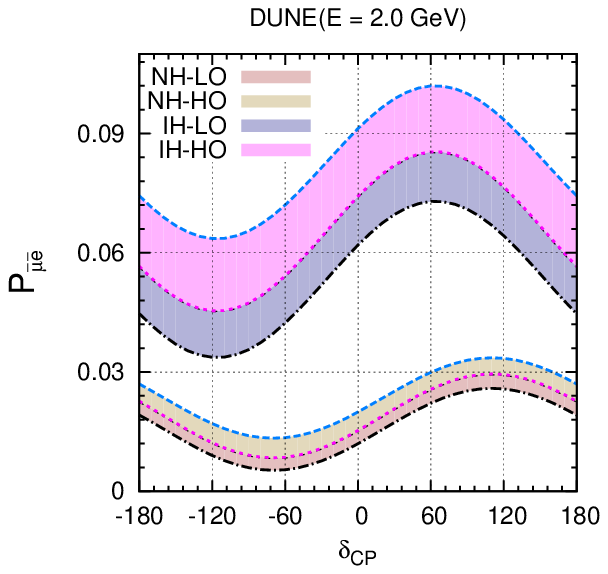}
        \end{tabular}
\vspace{-0.7cm}        
\caption{Left panel (right panel) 
represents $ P_{\mu e} (P_{\overline{\mu} \overline{e}}) $  for DUNE. 
Here the bands are over current 3$ \sigma $ range of $\theta_{23}$ ~\cite{global_valle}.  
For LO, NH (LO, IH) we consider  the range of $\theta_{23}$ over $ 38.8^\circ-45^\circ $( $ 39.4^\circ-55^\circ $) and 
for HO, NH (HO, NH) we consider  the range of $\theta_{23}$ over $45^\circ-53.3^\circ$($45^\circ-53.1^\circ$).}
\label{oct_prob_deg}
\end{figure*}

\begin{table*}
\begin{tabular}{| c | c | c | }
\hline
Octant Degeneracy & $ \nu $ & $\overline{ \nu } $ \\
\hline
 LHP, LO & degenerate with UHP, HO & no degeneracy  \\
 UHP, LO & no degeneracy & degenerate with LHP,HO \\
 LHP, HO & no degeneracy & degenerate with UHP, LO\\
 UHP, HO & degenerate with LHP,LO & no degeneracy \\
 \hline
\end{tabular}
\caption {The octant degenerate parameter space for neutrinos and antineutrinos.
Here, LO=Lower octant, HO=Higher octant, UHP=Upper half plane ($0^\circ < \dcp < 180^\circ$) and LHP=Lower half plane ($-180^\circ < \dcp < 0^\circ$).}
\label{table:octantdeg}
\end{table*}


The probabilities that are relevant for the DUNE experiment are $P_{\mu e}$ and
$P_{\mu \mu}$ and the corresponding probabilities for the antineutrinos. 
In presence of matter, the relevant oscillation probabilities can be  
expanded perturbatively in terms of small parameters
$ \alpha(\equiv \Delta m^{2}_{21}/ \Delta m^{2}_{31}) $ and $ \theta_{13} $ as
follows,~\cite{akhmedov,cervera,freund}
\begin{widetext}
\begin{eqnarray}
\label{p_mu_e}
P_{\mu e} &=&\underbrace{4 s^{2}_{13}s^{2}_{23}\frac{\sin^{2} (A-1)\Delta}{(A-1)^2}}_{\mathcal{O}_{o}}  
    +\underbrace{ \alpha^{2} \cos^{2}\theta_{23} \sin^{2}2 \theta_{12} \frac{\sin^{2} A\Delta}{A^2}}_{\mathcal{O}_{2}} \\ \nonumber 
      &+& \underbrace{\alpha s_{13} \sin 2\theta_{12}  \sin 2\theta_{23}\cos(\Delta+\delta_{cp}) \frac{\sin (A-1)\Delta}{(A-1)}\frac{\sin A\Delta}{A}}_{\mathcal{O}_{1}} \\ 
P_{\mu\mu} &=& 1 - \sin^2 2\theta_{23} \sin^2\Delta + \mathcal{O}( \alpha , s_{13}) \label{p_mu_mu} 
\end{eqnarray}
where, 
\begin{eqnarray}
  \Delta &\equiv& \frac{\Delta m^{2}_{31} L}{4E}, 
 A \equiv \frac{2EV}{\Delta m^{2}_{31} } \equiv \frac{VL}{2 \Delta}, 
{\rm and} ~ V = \pm \sqrt{2} G_F n_e 
\end{eqnarray}
\end{widetext}
These expressions are derived assuming constant  matter density approximation.
Similar expressions for antineutrino  probabilities can be obtained by 
replacing 
$\delta_{CP} \rightarrow - \delta_{CP}  $ and $ V \rightarrow  - V $.
The '$ + $($ - $)' sign here represents neutrino (antineutrino).  For NH, 
$ \Delta m^{2}_{31} $ is positive and for IH,
$ \Delta m^{2}_{31} $ is negative. 
Hence, in the neutrino oscillation probability $A$ is positive for 
NH and negative for IH. For antineutrinos, the sign of $A$ gets reversed.

It is clear from the above expressions that to leading order 
$P_{\mu \mu}$ suffers from intrinsic octant degeneracy 
between $\theta_{23}$ and $\pi/2 - \theta_{23}$. 
$P_{\mu e}$ does not suffer from intrinsic degeneracy
and the octant sensitivity comes mainly from this channel. 
However since $P_{\mu e}$ depends on $\sin^2\theta_{23}$, the 
$\chi^2$ is an increasing function of $\theta_{23}$ for this case
and the wrong octant minima from this channel always occurs for $45^\circ$. 
On the other hand $P_{\mu \mu}$ forces the minima to
$\sim \pi/2- \theta_{23}$, where the appearance channel has a large 
octant sensitive contribution. 

However although $P_{\mu e}$ does not suffer from intrinsic degeneracy 
it is possible to have 
\begin{equation} 
P_{\mu e}(\Delta, \theta_{23}^{tr}, \delta_{CP}^{tr})
= P_{\mu e}( \Delta,\theta_{23}^{wr},\delta_{CP}^{wr}),
\label{eq:rh-wo-wcp}
\end{equation}
 where the suffix ${tr}$ (${wr}$) denotes the true (wrong) values  of 
the parameters. 
The above equation implies  that apart from the true solution 
one can also get duplicate solutions with 
right hierarchy - wrong octant - wrong $\dcp$ (RH - WO - W$\dcp$).  
Note that unlike in the case of $P_{\mu \mu}$, for $P_{\mu e}$ 
one needs to consider the variation of $\theta_{23}$ over the 
whole of the opposite octant in order to identify the degenerate 
solution.   
Apart from this, if the hierarchy is unknown then one can also have 
\begin{equation} 
P_{\mu e}(\Delta,\theta_{23}^{tr},\delta_{CP}^{tr}) = P_{\mu e}(-\Delta, \theta_{23}^{wr}, \delta_{CP}^{wr}).
\label{eq:wh-wo-wcp} 
\end{equation} 
This corresponds to solutions with wrong hierarchy - wrong octant - wrong $\dcp$ (WH - WO - W$\dcp$).
As pointed out in \cite{Ghosh:2015ena} the most generalized case assuming 
$\theta_{13}$ as fixed, 
gives rise to total eight possibilities corresponding to different 
combinations 
of right (wrong) hierarchy and/or octant and/or $\dcp$.  
From Fig.~\ref{oct_prob_deg} one can see that for the DUNE baseline 
degenerate solutions with right-$\dcp$ do not come unlike the 
case of the experiments T2K and \nova \cite{Ghosh:2015ena}. 
This is because due to matter effects the bands for NH and IH 
are much more well separated and hence the intersection 
at right $\dcp$ do not occur. In this work we show how the octant sensitivity is affected by 
the wrong solutions defined in Eqs. \ref{eq:rh-wo-wcp} and \ref{eq:wh-wo-wcp}. We also discuss how the $\dcp$ sensitivity is affected by the occurrence of wrong octant solutions. We put emphasis on the role of antineutrinos and point out some unexpected behaviour due to matter effects.


Fig. \ref{oct_prob_deg} describes the oscillation probability in presence 
of earth matter for $\mathrm{L=1300~km}$
and $E = 2~{\rm GeV}$. 
The bands are due to the variation of $\theta_{23}$ (see figure caption for details).
The neutrino oscillation probability for NH gets significant enhancement 
in presence of  earth's matter 
as compared to IH as shown in the left panel.  It is seen that the maximum 
probability for NH
can become more than 3-times than that of IH. 
But in the case of antineutrinos the  scenario gets reversed 
as $A$ and
$\delta_{CP}$ changes their sign,  
as can be observed in  the right panel. 
This can be understood from Eq. \ref{p_mu_e}
that the $ {\mathcal{O}_{o}} $ term is $ \Delta $ dependent which 
enhances the probability value for 
the given set of oscillation parameters for NH as compared to IH for neutrino 
and ($ {\mathcal{O}_{1}}, {\mathcal{O}_{2}} $) terms are $ \alpha $ and $ \alpha^2 $ suppressed respectively.

Note that for vacuum oscillation maxima, $\Delta$ corresponds to $90^\circ$. 
Thus in the appearance channel probability (cf. Eq. \ref{p_mu_e}), 
$\dcp=-90^\circ$($+90^\circ$) 
correspond to maximum (minimum) point in the probability  for neutrinos.
For antineutrinos it is the opposite.  
Thus, for these values of $\dcp$, octant sensitivity is  expected to be 
maximum  if there is no degeneracy.  
Note that with the inclusion of matter effect, 
the appearance channel probability maxima does not coincide with the vacuum 
maxima and in that case the maximum and minimum
points in the probability  do not come exactly at $\pm 90^\circ$
but gets slightly shifted.  
This can be seen from Fig \ref{oct_prob_deg}. 
However for illustration, we will take $\dcp=\pm 90^\circ$ as the reference points to describe the 
physics of octant in DUNE.

It is to be observed  that, if we draw a horizontal line at particular 
probability value then the different intersection points with the given band lead to different degenerate solutions.  
The occurrence of octant degeneracies that can be inferred from these plots are summarized in Table \ref{table:octantdeg}.
From the above discussions as well as from earlier studies it is clear that the nature of octant - $\dcp$ degeneracy is 
different for neutrinos and antineutrinos and therefore combined neutrino-antineutrino run is helpful for resolving the octant degeneracy 
\cite{suprabhoctant, suprabhlbnelbno,Coloma:2014kca}.
Also note that the behaviour of octant-$\dcp$ degeneracy in neutrinos and antineutrinos   
is same for both NH and IH.

The probability plot as given in Fig.~\ref{oct_prob_deg} is done for an 
energy of 2 GeV. 
However it is possible that because of the broad-band nature of the 
beam the occurrence of degeneracy at a particular energy may not 
be true over the whole energy range. Thus for DUNE, one can still get some amount of octant sensitivity, even in 
the degenerate parameter space outlined in Table ~\ref{table:octantdeg},  
when integrated over all the energy bins. 

%
It is to be noted that Fig. \ref{oct_prob_deg} does not demonstrate any hierarchy degeneracy since the two bands corresponding to NH and IH remain non-overlapping.
However conclusions drawn at probability level need to be substantiated by a 
proper $\chi^2$ analysis to determine with what significance the hierarchy 
degeneracy is actually resolved by DUNE. Therefore we will present the results
of octant sensitivity either  for both cases -- right and wrong hierarchy or by marginalizing over the hierarchy.   
\section{Results }
\label{sec4}

\begin{figure*}
        \begin{tabular}{lr}
                \hspace*{-0.35in} 
                \includegraphics[width=0.5\textwidth]{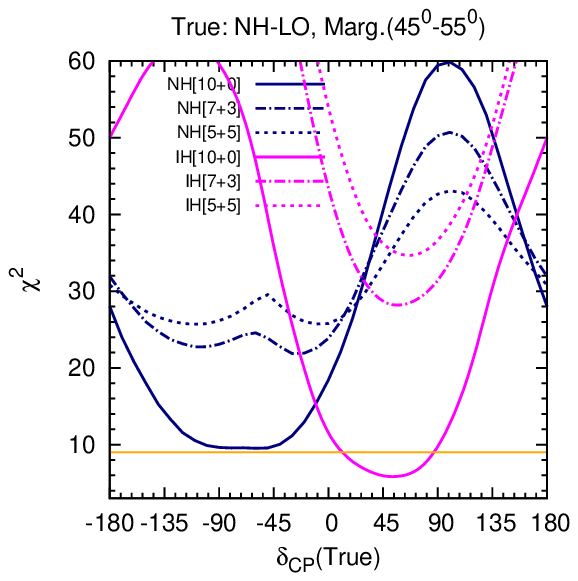}
                & 
                \hspace*{-1.3in}
                \includegraphics[width=0.5\textwidth]{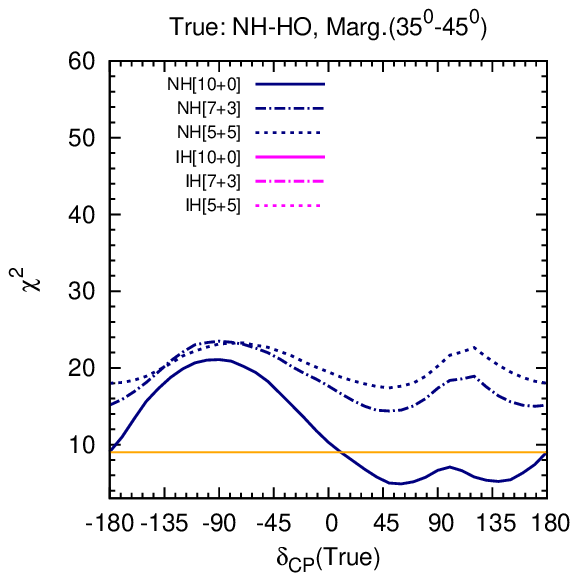}\\
                \hspace*{-0.35in} 
                \includegraphics[width=0.5\textwidth]{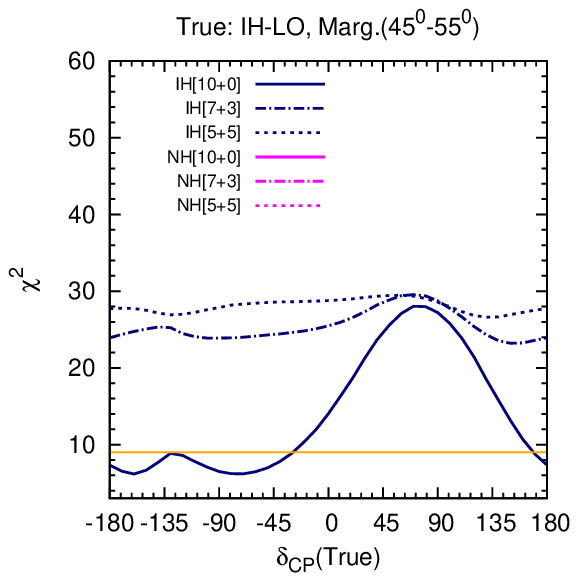}
                & 
                \hspace*{-1.3in}
                \includegraphics[width=0.5\textwidth]{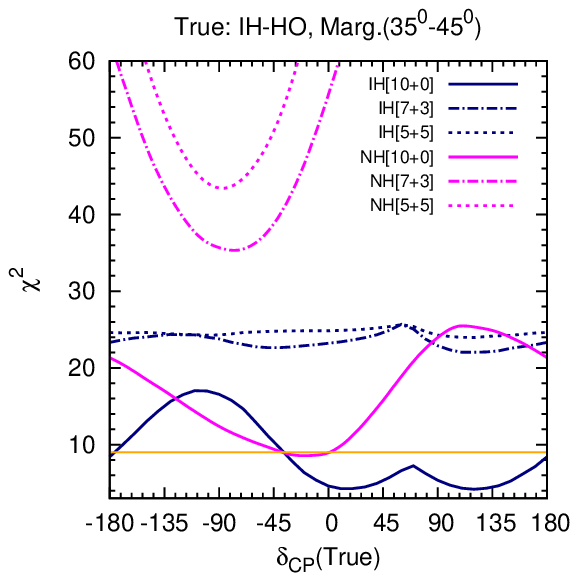}
        \end{tabular}
\vspace{-0.7cm}        
\caption{Octant discovery $ \chi^{2} $ for DUNE. Left (right) panel is for LO (HO), 
where true($ \theta_{23}$) is considered as $39^\circ$($51^\circ$) and test($ \theta_{23}$) is 
marginalized over ($ 45^\circ $ to $ 55^\circ $) for LO and ($ 35^\circ $ to $ 45^\circ $) for HO. The labels
NH, IH inside the plots signifies test hierarchy.}
\label{oct_discovery}
\end{figure*}
\begin{figure*}
        \begin{tabular}{lr}
                \hspace*{-0.35in} 
                \includegraphics[width=0.5\textwidth]{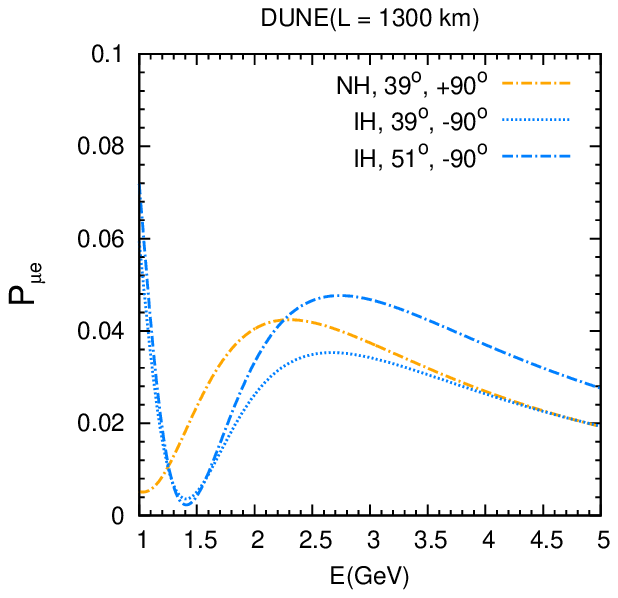}
                \hspace*{-1.0in}
                \includegraphics[width=0.5\textwidth]{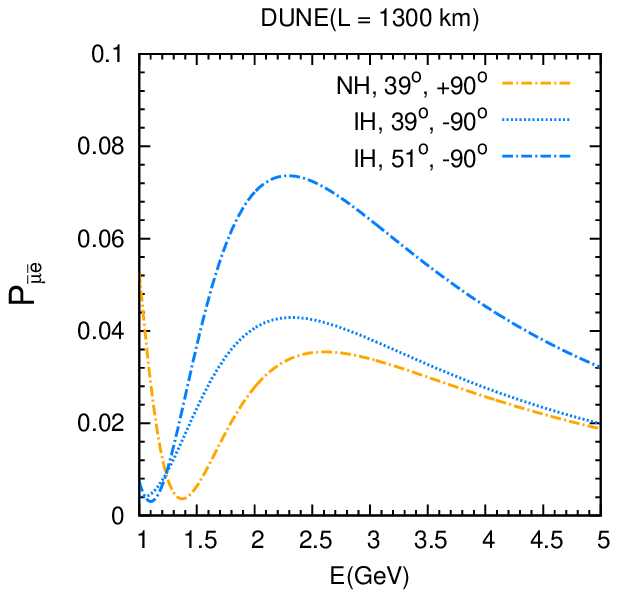}                     
        \end{tabular}
\vspace{-0.5cm}        
\caption{Here, left panel (right panel) represents $ P_{\mu e} (P_{\overline{\mu} \overline{e}}) $ 
as a function of energy  for DUNE and hierarchy corresponds to orange (light blue) curve is NH (IH).}
\label{broadband_39_51}
\end{figure*}
\subsection{Octant discovery $ \chi^{2} $ for a 10 kt  detector}
\label{sec4a}

\begin{figure*}
        \begin{tabular}{lr}
                \hspace*{-0.2in}
                \includegraphics[height=6cm,width=8cm]{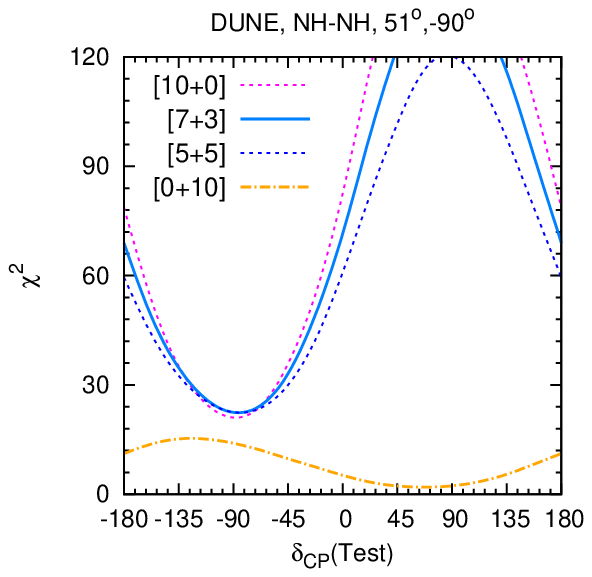}
                \hspace*{-1.2in}
                 \includegraphics[height=6cm,width=8cm]{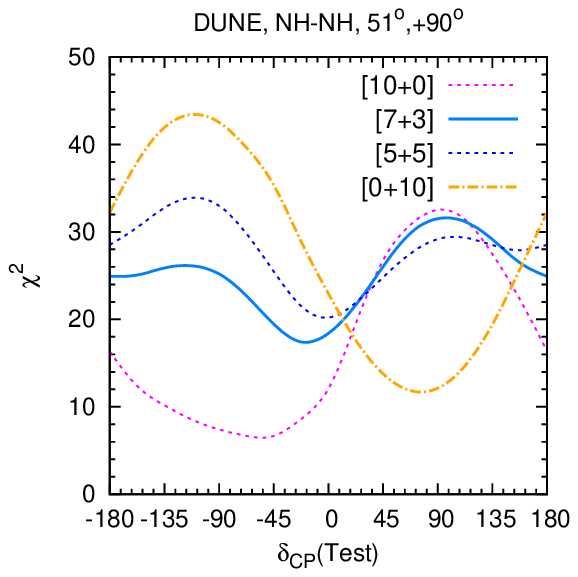}
                \hspace*{-1.2in}
                \includegraphics[height=6cm,width=8cm]{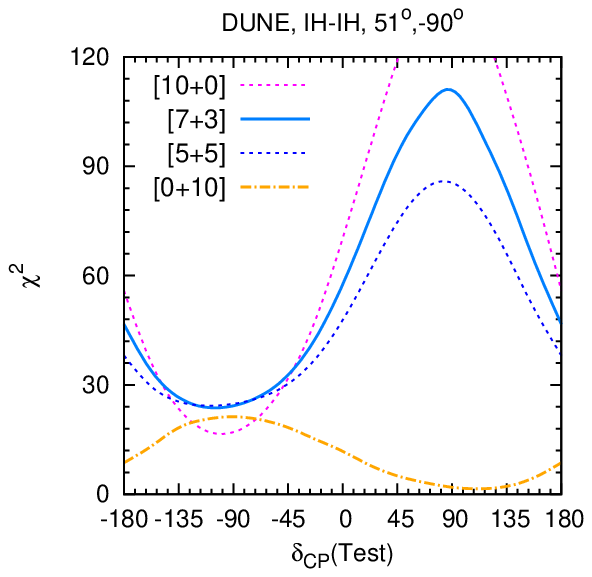}
        \end{tabular}
\vspace{-0.5cm}
\caption{Octant $ \chi^{2} $ vs test($ \delta_{CP} $) for DUNE.}
\label{fig:nuanu} 
\end{figure*}

In this section we discuss the octant sensitivity of DUNE for a 10 kt detector volume which is the projected detector volume for DUNE in the first phase.
The statistical $\chi^2$ for octant sensitivity is calculated by taking the correct octant in the true spectrum and 
the wrong octant in the test spectrum in the following formula 
\begin{eqnarray}
 \chi^2_{{\rm stat}} = \sum_i 2 \bigg[N_i^{{\rm test}} - N_i^{{\rm true}} - N_i^{{\rm true}} \log\bigg(\frac{N_i^{{\rm test}}}{N_i^{{\rm true}}}\bigg) \bigg],
\end{eqnarray}
where $N_i$ is the number of events in the $i^{\rm th}$ energy bin.
In Fig. \ref{oct_discovery} we show the $\chi^2$ for octant discovery which is the combined sensitivity coming from appearance channel, disappearance channel and 
$\sin^22\theta_{13}$ prior i.e.,
\begin{equation} 
\chi^2 = \chi^2_{ap} + \chi^2_{disap} + \chi^2_{prior}
\end{equation} 
as a function 
of true $\dcp$.    


We consider the  representative true values of $ \theta_{23} = 39^\circ$ for LO and $ \theta_{23} = 51^\circ$ for HO.
$\chi^2$ is marginalized 
over test values of $ \theta_{23} $ over opposite octant. 
We give the plots separately for true and false hierarchy. This shows for what parameters 
and to what extent the octant sensitivity is affected by the lack 
of knowledge of hierarchy.  
Depending on the true parameters, we get four combinations of 
(hierarchy$ - $octant): NH-LO, NH-HO, IH-LO, IH-HO. 
 For all the plots in the upper row of Fig. \ref{oct_discovery},  dark-blue curves are for True(NH)-Test(NH) and magenta
curves are for True(NH)-Test(IH) while for the lower row  dark-blue curves correspond to True(IH)-Test(IH) and magenta
curves correspond to True(IH)-Test(NH). 
Below we discuss the results for each true combination. 

\begin{itemize}
\item \textbf{NH-LO ($\theta_{23}^{\mathrm{true}} = 39^\circ$):}
The figure for true NH-LO shows that for  values of $\dcp$ in the 
lower half plane, a 10 year neutrino run of DUNE  
can resolve the octant degeneracy at 3$ \sigma $ C.L.
The inclusion of antineutrino run helps in enhancing the octant 
sensitivity  
for $\dcp$ in LHP ($-180^\circ < \dcp < 0^\circ$) and $\theta_{23}$ in LO since the  antineutrino probability is 
devoid of octant degeneracy.
Note that in this case though pure neutrino run suffers from octant degeneracy,
still we get $\chi^2$ around 10. This is one of the unique features of the 
broad-band beam where the degeneracy does not exist over the 
entire energy range and  one can still have  some octant sensitivity
from the neutrino channel. For the UHP ($0^\circ < \dcp < 180^\circ$) on the other hand the 
neutrino data gives a better octant sensitivity since 
antineutrinos are plagued with degeneracies
for LO, as shown by the blue curves. However the scenario changes if we assume the hierarchy 
is not known. In that case the antineutrino run is seen to help to remove 
wrong hierarchy-wrong octant solutions in-spite of having 
degeneracies, as is seen from the magenta curves. 
 In order to understand this point
we have plotted the appearance channel probability vs energy in Fig. \ref{broadband_39_51}.
The left panel is for neutrinos and the right panel is for antineutrinos. In the left panel we see that
the orange curve($\dcp=+90^\circ$) is well separated from the dotted blue curve($\dcp=-90^\circ$) near the oscillation maxima for $\theta_{23}=39^\circ$. 
But when marginalized over
$\theta_{23}$, the dashed blue curve which corresponds to $\dcp=-90^\circ$ and $\theta_{23}=51^\circ$,
overlaps with the orange curve to give WH-WO-W$\dcp$ solution\footnote{Due to the presence of $P_{\mu \mu}$ channel,
the wrong octant minima comes around $\theta_{23}=51^\circ$ for true $\theta_{23}=39^\circ$.}. On the other hand in the right panel we see that 
due the marginalization of $\theta_{23}$ the dashed blue curve moves far away from the orange curve resolving the degeneracy.
Note that if we marginalize over hierarchy
then for UHP the minimum will come at the WH solution 
with only neutrino data and hence octant degeneracy is not resolved 
at $3\sigma$ for  $9^\circ < \dcp <90^\circ$  belonging to the UHP. 
However with 7+3 years run the octant 
degeneracy is resolved with a $\chi^2 > 25$ even without the 
knowledge of the true hierarchy for all values of $\dcp$. 
With 5+5 year run in most part of UHP 
the minima occurs with the RH solution. But for  $45^\circ < \dcp <115^\circ$, 
the WH minima is below the one with RH. 
%

\item \textbf{NH-HO ($\theta_{23} = 51^\circ$)} 
For this case 
from Fig. \ref{oct_prob_deg} it is seen 
that for ($ 51^\circ $, -90$ ^\circ $, NH)  no octant 
degeneracy prevails 
at the probability level for neutrinos whereas antineutrinos have octant 
degeneracy. Also, antineutrinos have less statistics. 
Thus we expect that only neutrino run should give a better sensitivity. 
But, we notice  from the top right figure of Fig. \ref{oct_discovery},
that  addition of antineutrino gives higher $ \chi^{2} $ value as compared to only neutrino mode [10+0]. 
In order to understand this feature in the first panel of Fig. \ref{fig:nuanu}
we plot the $\chi^2$ vs test $\dcp$. 

The curve for only antineutrinos indeed confirm the occurrence of 
degeneracies close to $\dcp \sim 90^\circ$.  
However at that point the neutrino $\chi^2$ is very high. 
Thus, when the neutrino and antineutrino data are combined the 
overall minima is governed by the neutrinos and so comes close 
to the true value of $\dcp = -90^\circ$. At this point both 
neutrinos and antineutrinos have octant sensitive contribution.  
This is shown in Table \ref{table:appdisapp_nh} where we illustrate 
the contributions from the  neutrinos and antineutrinos separately 
for the appearance channel.
It is evident that as we increase the 
antineutrino component the contribution from neutrino channel reduces 
whereas that from the antineutrino channel increases.  
Thus although the antineutrino channel has degeneracy the 
minima does not come at the point of degeneracy as it is governed
by the neutrinos. Even then the total 
$\chi^2(= \chi^2_{ap, \nu} + \chi^2_{ap, \overline{\nu}})$ 
from appearance channel 
(11.13, 10.19), corresponding to $[7+3]$ and $[5+5]$ respectively, 
is less than the pure neutrino  run. 
However, the total $\chi^2$ for the mixed run is higher.

To understand this point we  list the contribution from the disappearance
$\chi^2$ and it is seen that although for pure neutrino run the disappearance
channel does not have any octant sensitive contribution to the total $\chi^2$ 
for mixed runs this channel also provide some octant sensitivity. 
This arises because due to matter effects the neutrino and antineutrino
probabilities are different and hence the $\chi^2$ minima comes at 
different places.
%
\begin{table*}
\begin{tabular}{|c | c | c | c | c |c | c| }
\hline
 ($ \nu + \overline{\nu} $) & Test parameters& $ \chi^{2}_{ap,\nu} $&$ \chi^{2}_{ap,\overline{\nu}} $ &  $ \chi^{2}_{disap,(\nu+\overline{\nu})} $ & Prior & Total \\
\hline
%
\cline{2-7}
& \multicolumn{6}{|c|}{NH, 51$ ^0 $, -90$ ^0 $(true)} \\
\cline{2-7}
\hline
[10+0](8+0) &$ \theta_{23} $=41.5$ ^{0} $(41$ ^{0} $) & 11.5(6.65) & 0 & 0.5(0.35) & 9(1.44) & 21.05(8.44) \\
 & $ \sin^{2}2\theta_{13} =0.115(0.106)$& & & & &\\
 \hline
[7+3](5+3) & same as [10+0](8+0)& 9.14(4.28) & 1.99(1.44) & 1.97(0.37) & 9(1.44) & 22.46(7.18)\\
\hline
[5+5](4+4) &same as [10+0](8+0) & 7.21(3.46) & 2.98(1.44) & 3.34(0.37) & 9(1.44) & 22.52(6.72)\\
\hline 
\cline{2-7}
& \multicolumn{6}{|c|}{IH, 51$ ^0 $, -90$ ^0 $(true)} \\
\cline{1-7}
[10+0](8+0) &$ \theta_{23} $=40$ ^{0} $(41$ ^{0} $),$ \delta_{CP}=- 105^{0} (-90^{0})$ & 10.86(5.23) & 0 & 0.09(1.47) & 5.76(1.44) & 16.71(8.14) \\
 & $ \sin^{2}2\theta_{13} =0.112(0.106)$& & & & &\\
 \hline
[7+3](5+3) & same as [10+0](8+0)& 8.22(3.36) & 8.10(1.33) & 1.62(0.96) & 5.76(1.44) & 23.71(7.09)\\
\hline
[5+5](4+4) &$ \theta_{23} $=40.5$ ^{0}(41 ^{0}) $,$ \delta_{CP}=- 120^{0}(-90^{0}) $ & 6.46(3.37) & 9.78(2.08) & 2.14(0.84) & 5.76(0.36) & 24.15(6.66)\\
& $ \sin^{2}2\theta_{13} =0.112(0.103)$& & & & &\\
 \hline 
\end{tabular} 
\caption {Here, [10+0], [7+3] and [5+5] refers to ($ \nu + \overline{\nu} $) runs of DUNE , where as (8+0), (5+3) and (4+4) refers to ($ \nu + \overline{\nu} $) runs of T2K. 
The numbers in the parenthesis correspond to T2K. 
Also ``Test parameters" refer to the test values where $ \chi^{2} $ minimum appears and remaining oscillation parameters are same as true parameters.}
\label{table:appdisapp_nh}
\end{table*}
\begin{figure*}
\vspace{-1.5cm}
        \begin{tabular}{lr}
               \hspace*{0.65in}
                \includegraphics[width=0.5\textwidth]{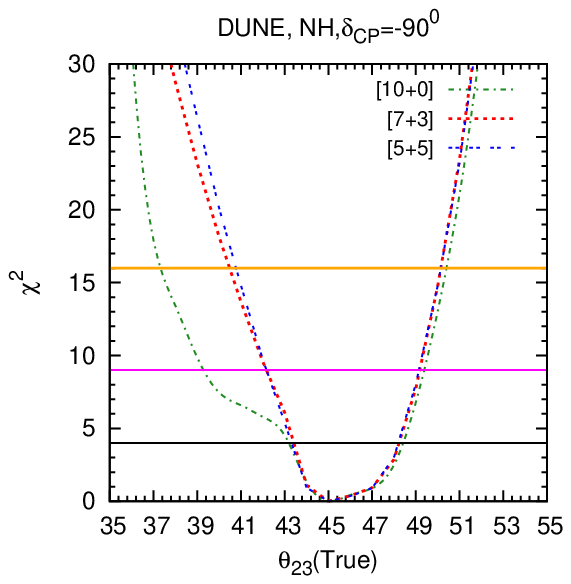}
                &
                \hspace*{-1.3in}
                \includegraphics[width=0.5\textwidth]{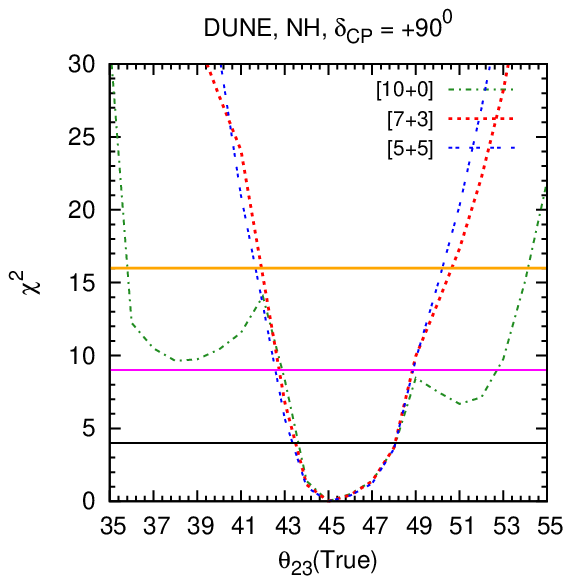}\\
                                \hspace*{0.65in}
                \includegraphics[width=0.5\textwidth]{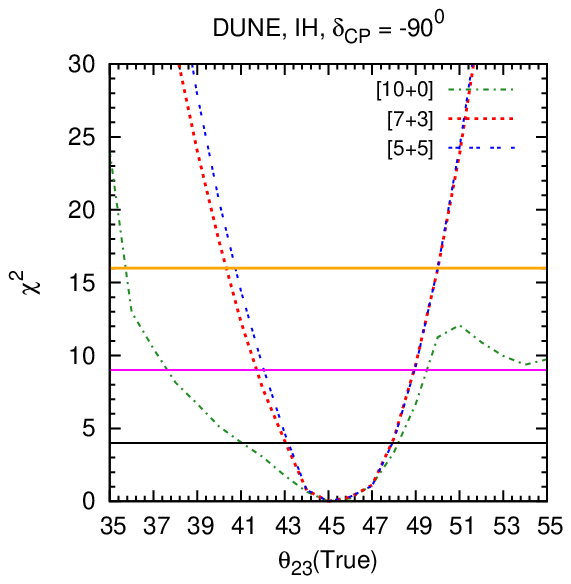}
                &
                \hspace*{-1.3in}
                \includegraphics[width=0.5\textwidth]{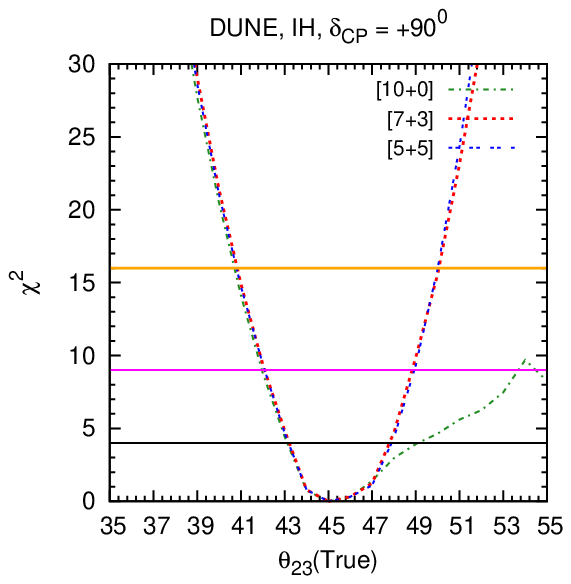}
        \end{tabular}
\vspace{1.3cm}
\caption{ Octant sensitivity $ \chi^{2} $ for DUNE. Left (right) panel is for $ \delta_{CP} = -90^\circ(+90^\circ) $, where true hierarchy is considered as NH(IH) for upper(lower) row. Here black, magenta and yellow lines represent $\chi^{2}$ value at  $2\sigma$, $3\sigma$ and $4\sigma$ respectively.}
\label{oct_sensitivity_10kt}
\end{figure*}
%
\begin{figure*}
        \begin{tabular}{lr}
                \hspace*{0.65in} 
                \includegraphics[width=0.5\textwidth]{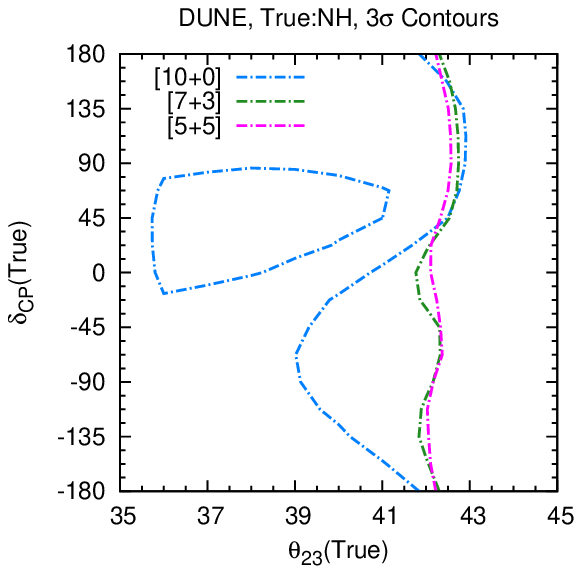}
                & 
                \hspace*{-1.3in}
                \includegraphics[width=0.5\textwidth]{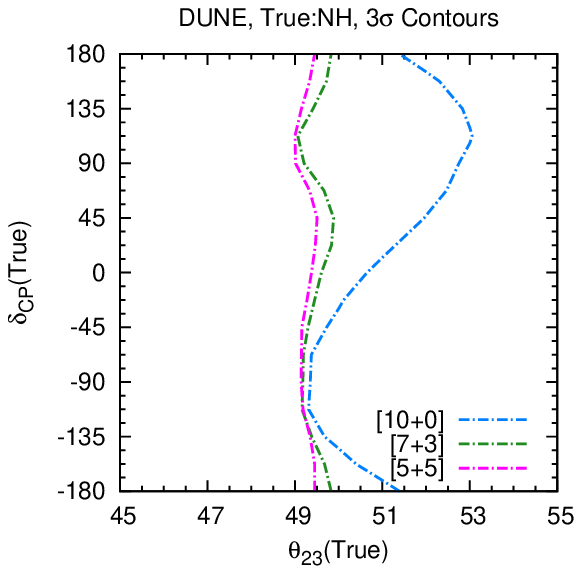}\\
                 \hspace*{0.65in} 
                \includegraphics[width=0.5\textwidth]{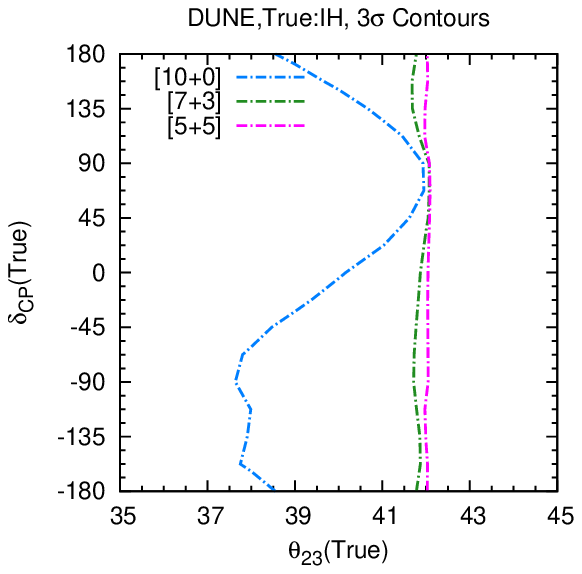}
                & 
                \hspace*{-1.3in}
                \includegraphics[width=0.5\textwidth]{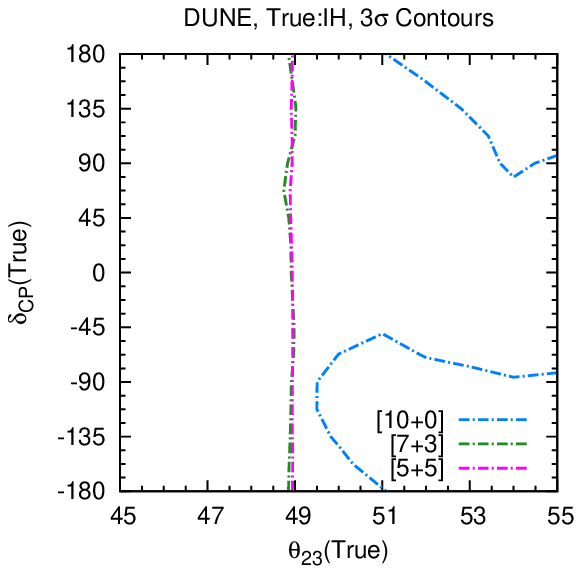}
        \end{tabular}
\vspace{-0.7cm}        
\caption{ Contour plots in true($  \theta_{23},\delta_{CP} $) 
plane, here true hierarchy is NH (IH) for upper (lower) row and left(right) panel is for LO (HO).  Marginalization over hierarchy is done.
The allowed regions are to the right (left) side of the contours in
the left (right) panel.}
\label{oct_lo_ho}
\end{figure*}
When one combines neutrino and antineutrino run 
then this creates a synergy and hence some octant sensitivity arises
from the disappearance channel also. 
Due to this reason when one combines appearance and disappearance channels
then addition of antineutrino runs actually gives a slight increase 
in $\chi^2$.
In the UHP on the other hand the octant sensitivity increases with antineutrino 
run. This is clear since for $P_{\mu e}$ the neutrino channel suffers from
octant degeneracy whereas the antineutrino channel does not and the addition 
of antineutrinos help to overcome the degeneracy. 
To illustrate this point further in the middle panel of Fig. \ref{fig:nuanu}
we plot the $\chi^2$ vs test $\dcp$ for true values ($51^\circ$, $90^\circ$). 
In this case the pure neutrino run gives the minima in the LHP
close to $\dcp \sim -45^\circ$ whereas pure antineutrino gives minima near the 
true value. However when we combine neutrino and antineutrino runs then the 
overall minima comes in between and moves towards the antineutrino minima 
as the $\bar{\nu}$ component is increased. At this point there is 
octant sensitive contribution from both neutrinos and antineutrinos. 
Thus the antineutrino data helps in this case by trying to shift the 
minima away from the degenerate point.
We also compare the $\chi^2$ for DUNE with that of T2K, 
given in parentheses in Table ~\ref{table:appdisapp_nh}, to understand the 
role of broadband beam and enhanced matter effect. It is seen from the last 
column that for T2K the $\chi^2$ reduces with increasing antineutrinos 
as is expected. Note that this is in contrast to DUNE due to its broadband nature and enhanced matter effect.

\item \textbf{IH-LO ($\theta_{23} = 39^\circ$) :} 
In this case for LHP the antineutrino run enhances the sensitivity 
because they  do not suffer from octant degeneracy 
as can be seen from Table. \ref{table:octantdeg}. 
But for the UHP the antineutrino probability has octant degeneracy. 
Thus again we expect that in UHP adding antineutrino data should 
reduce the sensitivity. But the figure shows a slight enhancement. 
This can again be explained by similar reasoning as for the 
NH, $51^\circ$ and $-90^\circ$ case. 
There is also the finite contribution from the disappearance channel 
enhancing the octant sensitivity 
when the neutrino and antineutrino runs are combined.
These combinations of hierarchy$ - $octant can resolve octant degeneracy at 
5$ \sigma $ C.L. with [5+5] years of [$ \nu + \overline{\nu} $] run for any 
value of true $ \delta_{CP} $ as shown in Fig. \ref{oct_discovery}.  

\item \textbf{IH-HO:($\theta_{23} = 51^\circ$)} 
For this case, for $\dcp$ in LHP the octant sensitivity with pure neutrino run 
is seen to be above $\chi^2=9$ in the interval $-180^\circ < \dcp < -45^\circ$. 
Adding antineutrino data helps to raise the $\chi^2$ for octant sensitivity. 
As before we ask the question how antineutrino data is helpful despite 
the presence of degeneracies in this channel. 
This can be explained again similar to the NH-HO case. The third panel 
of Fig. \ref{fig:nuanu} shows that for pure antineutrinos, 
there is very small octant sensitivity and the minima comes 
in the UHP between $90^\circ$ and $135^\circ$. 
However at the point, in the LHP, where the pure neutrino $\chi^2$ is minimum,
antineutrino $\chi^2$ has a large non-zero value and for combined
runs the minima is still governed by the neutrinos. Thus the contributions from the antineutrinos are also being added up in-spite of having degeneracy. 
The neutrino and antineutrino contributions from the 
appearance channel are shown in Table \ref{table:appdisapp_nh}. 
It is seen that for IH, because of the 
enhancement of the antineutrino probability due to  matter effect,  
a large octant sensitive contribution to the   
$\chi^2$ is obtained. The disappearance channel also gives a small contribution but the 
contribution from the antineutrino channel is almost comparable or larger
than the neutrino channel. 
It is also to be noted that if hierarchy is not known then for some
values of $\dcp$ the minima comes in the wrong hierarchy region for pure neutrino
run and the sensitivity is further reduced. Addition of antineutrinos resolves the 
hierarchy with $\chi^2 \geq 25$ and so the minima does not occur anymore for 
wrong hierarchy solution. For the UHP the only neutrino run has very poor sensitivity
due to degeneracies with $\dcp$ and addition of antineutrino runs help. 
The  UHP is more favourable for resolution of hierarchy-$\dcp$ degeneracy 
and even with only neutrino run hierarchy is resolved at $3\sigma$ for all values
of $\dcp$. 
Overall, close to $\chi^2=25$ sensitivity is achieved for this combination of hierarchy 
and $\theta_{23}$ with 7+3 or 5+5 combination for the whole range of $\dcp$. 
For this case also in Table ~\ref{table:appdisapp_nh} the T2K $\chi^2$ values 
are given in parentheses. It is seen from the last column that the 
overall $\chi^2$ for T2K decreases with enhanced antineutrinos unlike that
in DUNE. If one compares the appearance  $\chi^2$ values for the antineutrino
channel for DUNE and T2K  then it is seen that the contribution of this channel for DUNE is 
quite high and comparable or even greater than the neutrino contribution. 
This is due to the enhanced matter effect associated with IH and HO for the 
longer baseline of DUNE.

\end{itemize}
After discussing the role of antineutrinos and disappearance channel in octant sensitivity for DUNE,
in Fig. \ref{oct_sensitivity_10kt} 
we present the octant $\chi^2$ as a function of true $\theta_{23}$ 
for maximal CP violation. 
Depending on if the  true hierarchy is NH or IH and true $\dcp$ is 
$ \pm 90^\circ $ we get 4 possible combinations.
From these figures one can read off the range of $\theta_{23}$ for which 
octant can be determined for $\dcp =\pm 90^\circ$ at a specified C.L. 
We see for all the four cases of Fig.~\ref{oct_sensitivity_10kt}  
that with 7+3 years of ($\nu +\bar{\nu}$) run octant can be determined
at $3\sigma$($4\sigma$) 
for $\dcp = \pm 90^\circ$ excepting for the range 
$41.5^\circ < \theta_{23} < 49^\circ$($40.5^\circ < \theta_{23} < 50.7^\circ$).
From the figures we also see that 7+3 and 5+5 combinations give 
almost same sensitivity.
However for the pure neutrino run the ranges are different and  
also vary depending on the true values of $\dcp$ and hierarchy. 
In Table \ref{range10} we give the ranges of $\theta_{23}$ for 
which
octant can be resolved at $3\sigma$ and $4\sigma$ with pure neutrino run.
\begin{table*}
\begin{tabular}{| c | c | }
\hline
True Parameter & $\theta_{23}$ range for $3\sigma$($4\sigma$) \\
\hline
 NH, $\dcp=-90^\circ$ & $<39^\circ$($37.4^\circ$) and $> 49^\circ$($50.6^\circ$)  \\
 NH, $\dcp=90^\circ$ & $<43^\circ$($35.7$) and $> 53^\circ$($54^\circ$)  \\
 IH, $\dcp=-90^\circ$ & $<37^\circ$($35.7$) and $> 49^\circ$($55^\circ$)  \\
 IH, $\dcp=90^\circ$ & $<42^\circ$($40$) and $> 54^\circ$($55^\circ$)  \\
 \hline
\end{tabular}
\caption {Ranges of $\theta_{23}$ for which octant can be resolved at $3\sigma$ ($4\sigma$) for [10+0] configuration for 10 kt detector.}
\label{range10}
\end{table*}

So far we have focused on the cases for which either true $\theta_{23}$ was fixed or true $\dcp$ was fixed. 
In  Fig.\ref{oct_lo_ho} 
we give the 3$ \sigma $ exclusion plots in true($ \theta_{23}- \delta_{CP}$) plane. 
We consider all possible true values of $ \delta_{CP} $ from ($ -180^\circ $ to $ + 180^\circ$) and $\theta_{23}$ in lower octant from $35^\circ - 45^\circ$ and 
higher octant from $45^\circ - 55^\circ$. 
This figure shows the role of antineutrino run  in the full range of allowed
$\dcp$ and $\theta_{23}$ parameter space. The allowed region for the left (right) panel is 
the R.H.S (L.H.S) of each curve of the true($ \theta_{23}- \delta_{CP}$) plane \footnote{ For NH-LO, DUNE[10+0] (top left panel of Fig. \ref{oct_lo_ho}), the area
enclosed by the blue curve also corresponds to the allowed region.}.  
We observe by comparing the left and the right panels that
DUNE can provide better constraints on $ \theta_{23} $ parameter space
in case of LO as compared to HO. 
For NH-LO the antineutrino run is necessary for the LHP and part of UHP. 
Only in the range $90^\circ < \dcp < 135^\circ$ the only neutrino run i.e., the [10+0] configuration gives a slightly better sensitivity.   
On the other hand for NH-HO  the antineutrinos play a more prominent 
role for $\dcp$ in the UHP. 
For IH-LO the antineutrino run is again important apart from near 
$\dcp \sim 90^\circ$, for which the improvement in sensitivity by adding 
antineutrinos is not very significant.  
For IH-HO the antineutrinos play important role in the full parameter space. 
Also the exclusion plots show that if true 
$\theta_{23}$ lies between (43$^\circ - $49$^\circ$) then
it is not possible to resolve octant degeneracy at $3 \sigma$ C.L. by DUNE using 
10 kt detector. 
Overall one can say that antineutrino runs are necessary for 
most of the parameter region and 7+3 and 5+5 give similar sensitivities. Note that in the context of LBNO 75\% - 25\% ($ \nu - \overline{\nu} $) was recommended in \cite{Das:2014fja}.

\begin{figure*} 
        \begin{tabular}{lr}
                \includegraphics[width=0.56\textwidth]{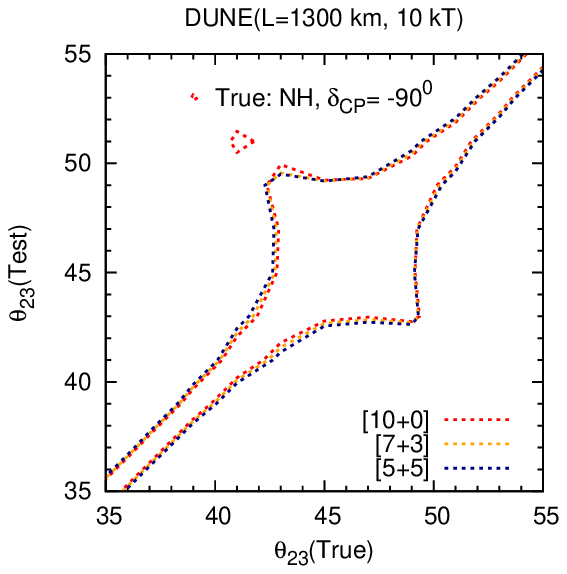}
                \hspace*{-1.0in}
                 \includegraphics[width=0.56\textwidth]{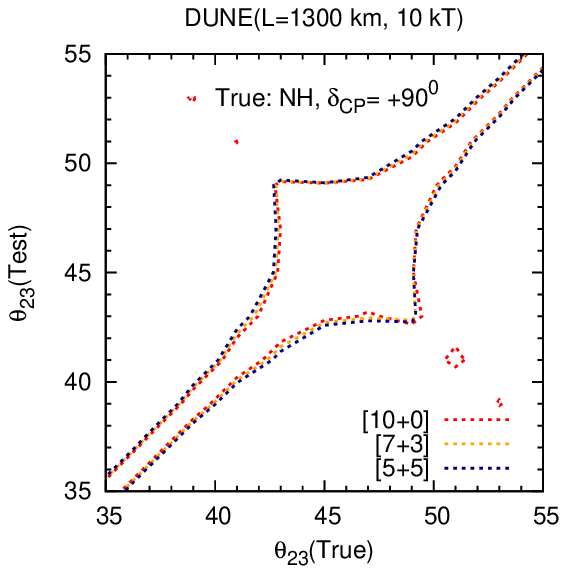} \\
                \includegraphics[width=0.56\textwidth]{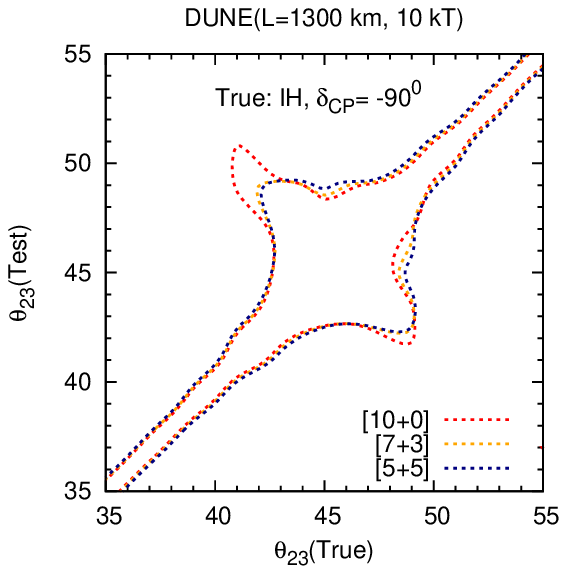}              
                 \hspace*{-1.0in}
                \includegraphics[width=0.56\textwidth]{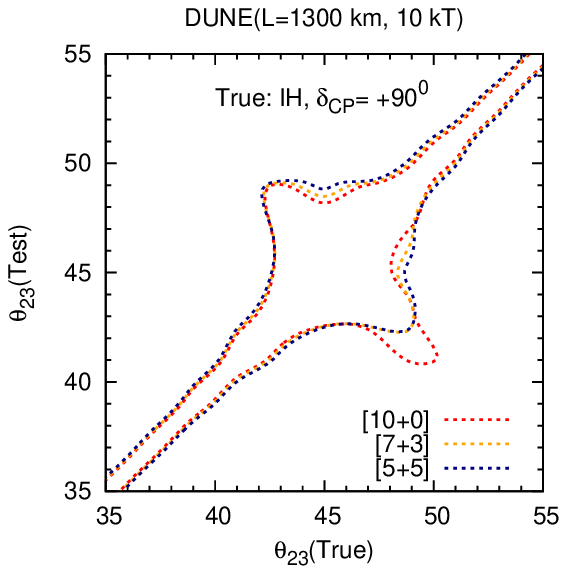}              
        \end{tabular}
\vspace{0.9cm}
\caption{$ \theta_{23} $ precision plots of DUNE in True($ \theta_{23} $) -  Test($ \theta_{23} $) plane at 3$ \sigma $  C. L.  Here top(bottom) row is for NH(IH).}
\label{pre} 
\end{figure*}

Finally in Fig. \ref{pre} we plot the $3 \sigma$ precision contours in the true $\theta_{23}$-test $\theta_{23}$ plane for $\dcp=\pm 90^\circ$. These figures reflect 
the relation between octant degeneracy and precision of $\theta_{23}$. The upper panels are for normal hierarchy and the lower panels are for 
inverted hierarchy. From these plots we see that for pure neutrino run there are other allowed values of $\theta_{23}$ apart from the true value, if $\theta_{23} \in$ LO (HO) at $\dcp=-90^\circ (+90^\circ)$ . 
This happens because of the octant degeneracy. 
As we have already seen, for $\dcp=-90^\circ (+90^\circ)$, neutrinos suffer 
from octant degeneracy in LO (HO) in both the hierarchies and this in turn
affects the precision of $\theta_{23}$ which is clearly seen
from the figures. Adding antineutrinos help to improve the precision and
both 7+3 and 5+5 give almost similar precision of $\theta_{23}$. 
But as one approaches the maximal value of $\theta_{23}$, the precision becomes worse due to the difficulty in determining the octant around those values of $\theta_{23}$.
\subsection{Antineutrinos, Octant Degeneracy and CP discovery potential of DUNE }
\label{sec4b}

In this section we present the CP discovery $\chi^2$ of DUNE as 
a function of true $\dcp$. CP violation discovery potential of an experiment is defined by its capability of distinguishing a true value of $\dcp$ other than $0^\circ$ and $180^\circ$. 
We present these figures for the case 
where hierarchy and octant are assumed to be unknown and known. 
The main aim of this section is to elucidate the role of antineutrinos
in discovering $\dcp$ and the interconnection  with  the octant 
degeneracy. 

\begin{figure*}
        \begin{tabular}{lr}
               \hspace*{0.65in}
                \includegraphics[width=0.5\textwidth]{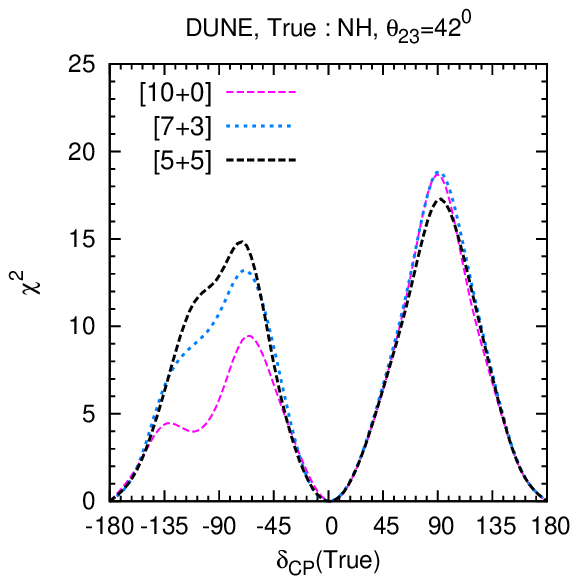}
                &
                \hspace*{-1.3in}
                \includegraphics[width=0.5\textwidth]{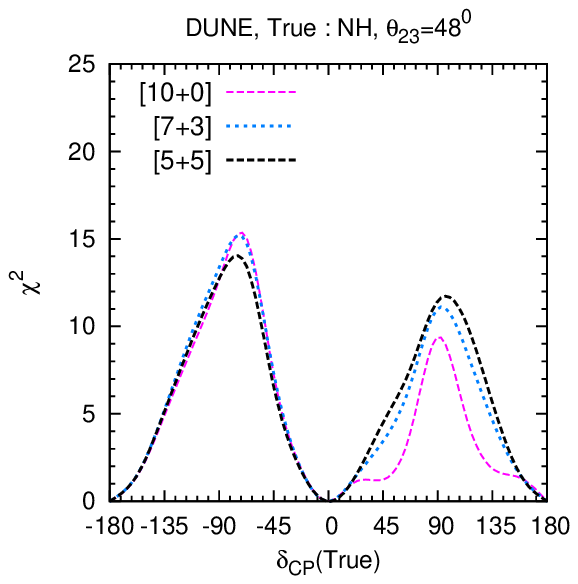}\\
                                \hspace*{0.65in}
                \includegraphics[width=0.5\textwidth]{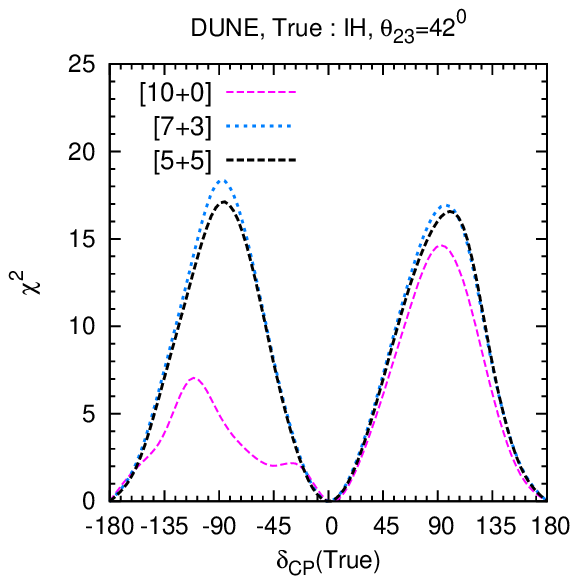}
                &
                \hspace*{-1.3in}
                \includegraphics[width=0.5\textwidth]{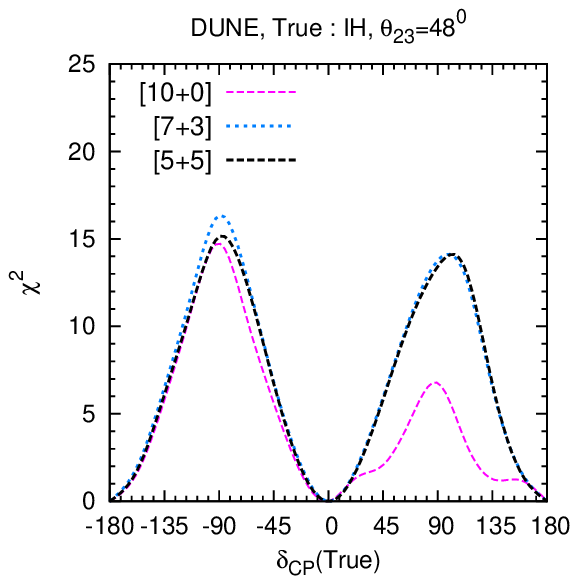}
        \end{tabular}
\vspace{-0.7cm}
\caption{ CPV $ \chi^{2} $ for DUNE when hierarchy and octant are unknown.}
\label{dune_cp_1}
\end{figure*}


\begin{figure*}
        \begin{tabular}{lr}
               \hspace*{0.65in}
                \includegraphics[width=0.5\textwidth]{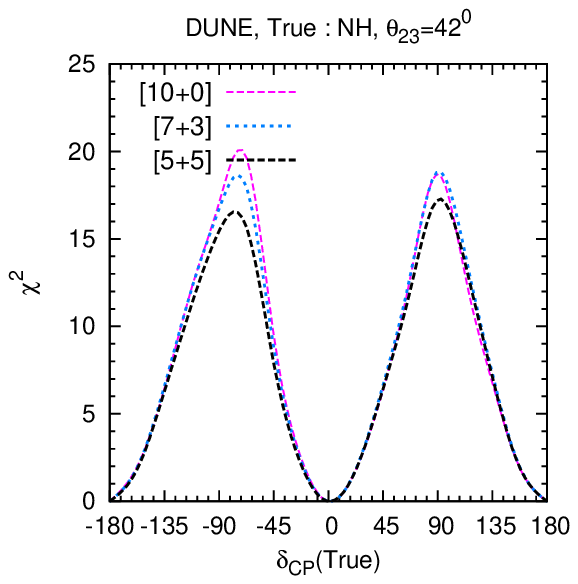}
                &
                \hspace*{-1.3in}
                \includegraphics[width=0.5\textwidth]{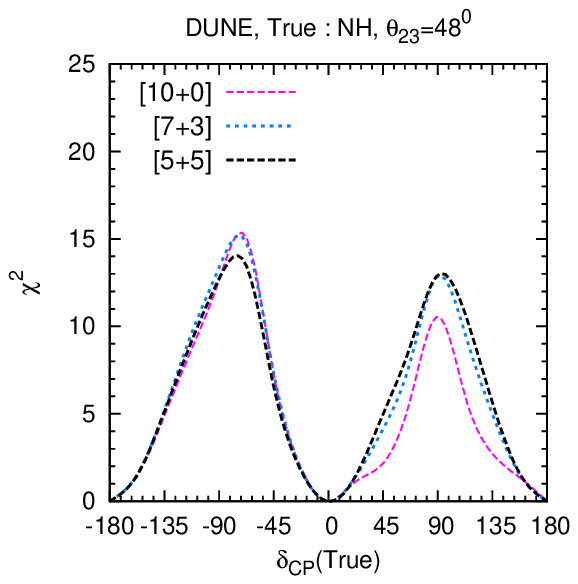}\\
                                \hspace*{0.65in}
                \includegraphics[width=0.5\textwidth]{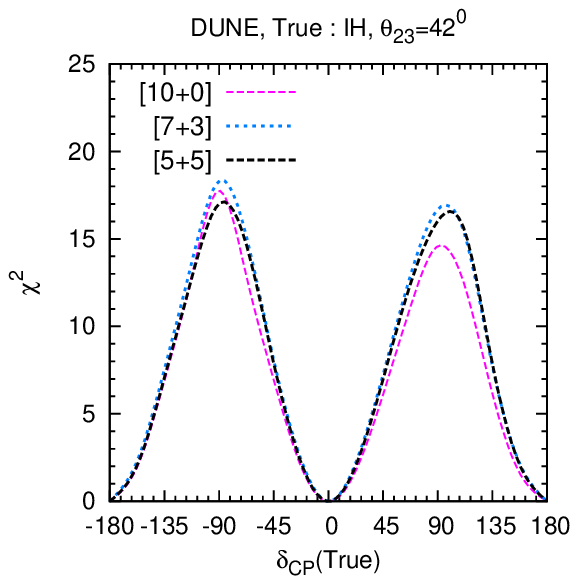}
                &
                \hspace*{-1.3in}
                \includegraphics[width=0.5\textwidth]{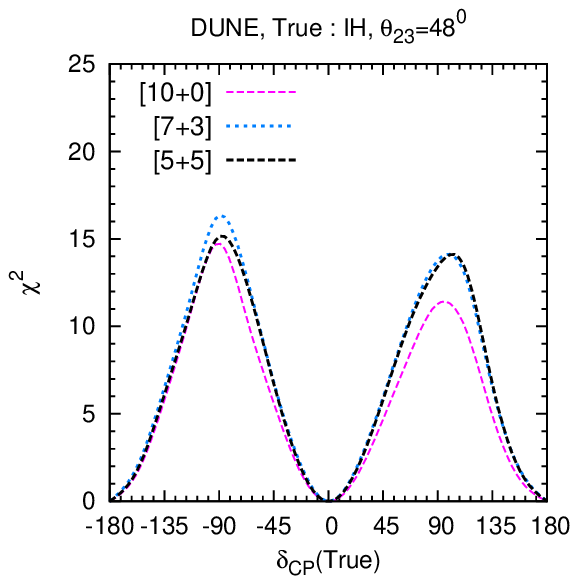}
        \end{tabular}
\vspace{-0.7cm}
\caption{ CPV $ \chi^{2} $ for DUNE when hierarchy and octant are known.}
\label{dune_cp_2}
\end{figure*}

\begin{figure*}
        \begin{tabular}{lr}
         \hspace*{0.65in}
         \includegraphics[width=0.5\textwidth]{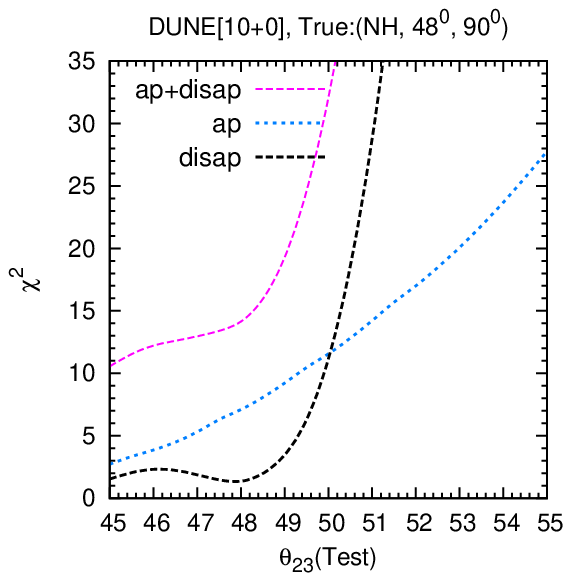}
         &
         \hspace*{-1.3in}
        \includegraphics[width=0.5\textwidth]{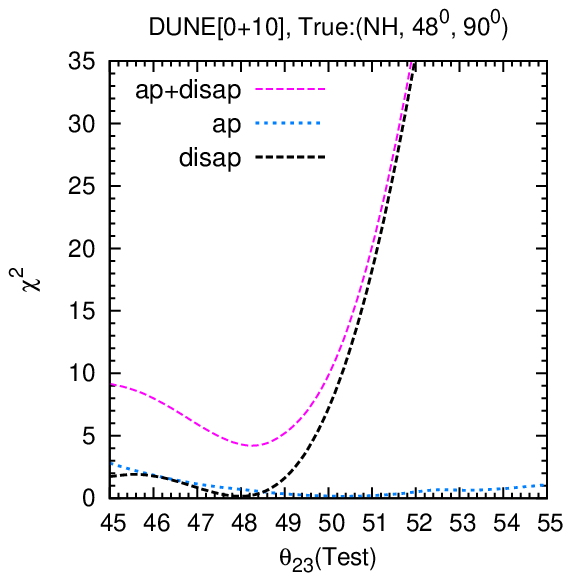}\\
         \hspace*{0.65in}
         \includegraphics[width=0.5\textwidth]{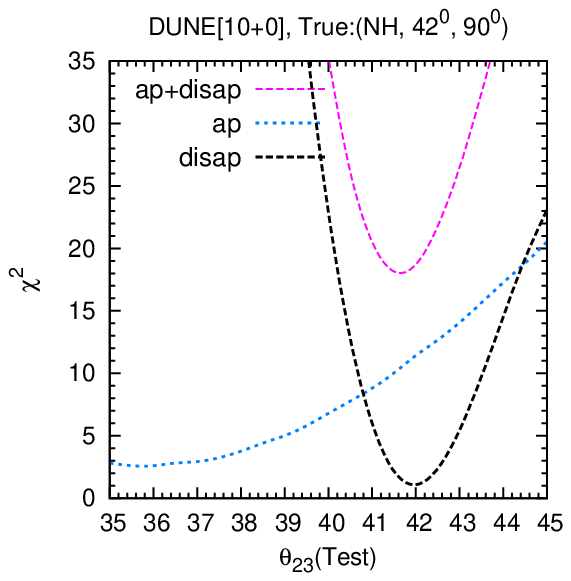}
         &
        \hspace*{-1.3in}
        \includegraphics[width=0.5\textwidth]{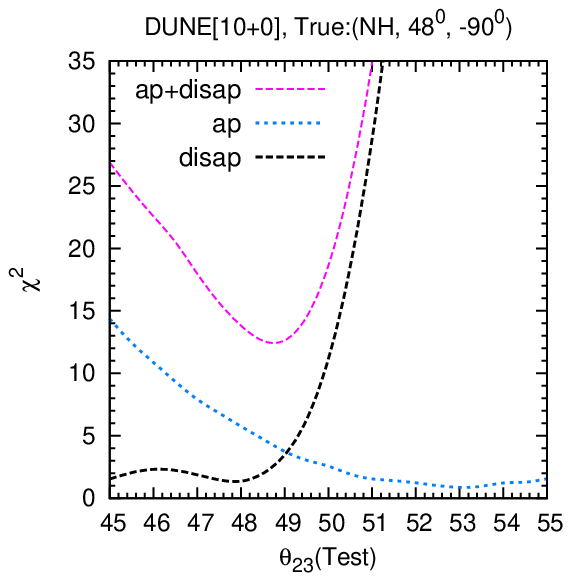}\\
        \hspace*{0.65in}
        \includegraphics[width=0.5\textwidth]{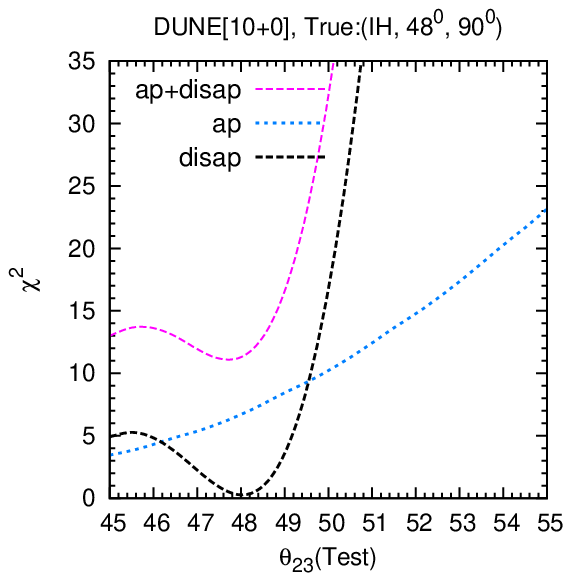}
        &
       \hspace*{-1.3in}
       \includegraphics[width=0.5\textwidth]{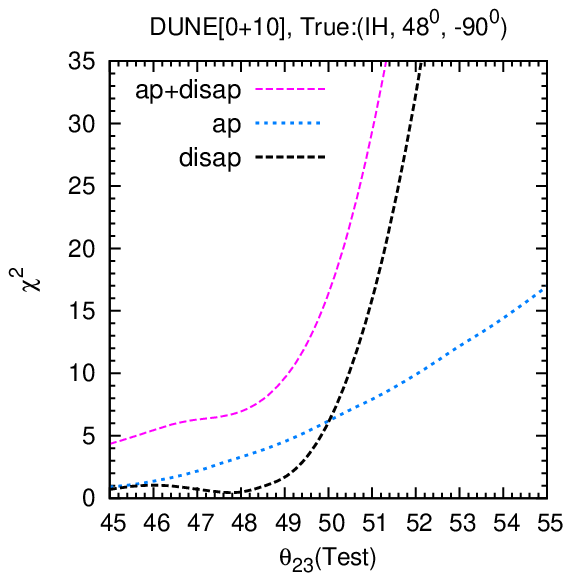}
        \end{tabular}
\vspace{-0.7cm}
\caption{ CPV $ \chi^{2} $ for DUNE when hierarchy and octant are known.}
\label{dune_cp_3}
\end{figure*}

The figure \ref{dune_cp_1} plots the CP discovery $\chi^2$ as function of 
true $\dcp$ for the case when hierarchy and octant are assumed to be unknown. 
From the different panels it is seen that:

\begin{itemize} 
\item 
The antineutrino runs play an
important role for (i) LO  near true $\dcp = -90^\circ$
and 
(ii) HO near true $\dcp = +90^\circ$. This is true for both NH and IH. 
Note from Table  \ref{table:octantdeg} that these are the regions where
neutrino probabilities exhibit octant degeneracy. Since antineutrino 
probabilities do not possess this degeneracy, addition of these helps in 
the removal of the degeneracy and enhancement of CP sensitivity. 

\item 
For true hierarchy as NH , +90$ ^\circ $-LO and -90$ ^\circ $-HO do not have 
octant degeneracy for neutrinos whereas antineutrinos have degeneracy 
(see Table \ref{table:octantdeg}). Even then 7+3  gives almost same result as 
10+0 notwithstanding the loss of statistics. In both cases this 
happens due to tension  between the neutrino and antineutrino 
$\chi^2$s.   

\begin{itemize} 
\item 
For +90$ ^\circ $-LO the minima for 10+0 comes at $\dcp = 180^\circ$ whereas replacing 
3 years 
of neutrino run by antineutrino run shifts the $\chi^2_{min}$ at $\dcp = 0^\circ$ 
where the neutrino contribution is higher and thus 7+3 becomes
comparable to 10+0.  

\item  For the case of -90$ ^\circ $-HO and neutrinos the  CPV $\chi^2$ is a falling function of $\theta_{13}$, and the minima comes at 0.109 while for 7+3 it comes at 0.106.
The neutrino contribution at $\sin^2\theta_{13} = 0.106$ being higher
the overall $\chi^2$ for 7+3 becomes greater.  
\end{itemize} 

\item 
Similarly,  for true hierarchy IH, +90$ ^\circ $-LO and -90$ ^\circ $-HO 
are free
from octant degeneracy for neutrinos.
But still the CP sensitivity for these cases
are slightly better for  combined neutrino-antineutrino run (for both 
7+3 and 5+5 case) as compared to pure neutrino run. This happens because  due to  matter effects 
the $P_{\overline{\mu} \overline{e}}$ is higher than $P_{\mu e}$ for IH (see Fig. \ref{oct_prob_deg}).  Thus  addition of  antineutrinos  enhances the appearance $\chi^2$.    

\end{itemize} 
%
 
In figure \ref{dune_cp_2} we present the same plots as that of 
Fig. \ref{dune_cp_1} but assuming the hierarchy and octant to be known.  

\begin{itemize} 
\item 
Comparing with the plots in Fig. \ref{dune_cp_1} we see that the CP sensitivity 
for -90$^\circ$-LO 
improves for the 10+0 case, for both the hierarchies.
In fact for  -90$^\circ$-LO-NH, 10+0 gives the 
best sensitivity if the octant is known. 
This establishes the fact that, 
the antineutrino run was instrumental for removing the wrong octant 
solutions.  

\item 
For +90$^\circ$-NH-HO although there is some improvement  
for the 10+0 case as compared to the case of unknown octant, 
the CP sensitivity of 7+3 and 5+5 are still better than 10+0. 
This implies that though octant is known, antineutrinos play some role in 
enhancing the CP sensitivity.

\begin{itemize}
\item
To understand the above point in more detail  in  Fig. \ref{dune_cp_3} 
we plot the CPV discovery $\chi^2$ for different channels vs test $\theta_{23}$ in the correct octant (since octant is assumed to be known) for a particular value of true $\dcp$. The top left panel of 
Fig. \ref{dune_cp_3} shows that, for neutrinos 
and true value of +90$^\circ$-NH-48$^\circ$, 
the CP sensitivity of the appearance channel is an increasing function of 
test $\theta_{23}$. But as the precision of $\theta_{23}$ (which comes from the disappearance channel) is poor near the maximal value, the combined $\chi^2$ minimum does not
occur at the true $\theta_{23}$ value ( which is $\theta_{23}=48^\circ$) but occurs at $\theta_{23}=45^\circ$.For antineutrinos (top right panel), the nature of the
disappearance channel  $\chi^2$ is same as that of neutrinos but  
the appearance  $\chi^2$ decreases as test $\theta_{23}$ increases. 
Because of this opposite behaviour when antineutrinos
are combined with neutrinos, the $\chi^2$ minima shifts to the correct value of $\theta_{23}$ and the overall CP sensitivity at the true point is enhanced. 
The poor $\theta_{23}$ precision of $+90^\circ$-HO-NH in
neutrinos arise due to the higher matter effect. 
Due to which the subleading matter terms start to contribute in the disappearance channel
which affect the $\theta_{23}$ precision. 
Note that this does not happen for $-90^\circ$-NH-HO (despite matter is high) because 
even though the precision of $\theta_{23}$ is poor near the maximal value of 
$\theta_{23}$ for $\dcp=-90^\circ$ the appearance channel sensitivity is a 
decreasing function of test $\theta_{23}$ and this causes the overall minima to occur at the 
correct value of $\theta_{23}$. This can be seen from the
right panel of the middle row of Fig. \ref{dune_cp_3}

\item 
Also note that for true $+90^\circ$-NH  
42$^\circ$  (left panel of middle row of Fig. \ref{dune_cp_3}),
the precision of $\theta_{23}$ near the maximal value is quite good as 
compared to $\theta_{23}=48^\circ$. This is because for lower octant, the denominator in the $\chi^2$ is smaller as compared to  that in HO and thus a 
better $\theta_{23}$ precision is obtained.  

\item
For true hierarchy IH, even $\theta_{23}=48^\circ$ has  a good precision 
near the maximal value for neutrinos (bottom left panel of Fig. \ref{dune_cp_3}).
This is because for IH the matter effect is less for neutrinos and thus
$\theta_{23}$ precision measurement capability of the disappearance channel is better as can 
be seen comparing the top and bottom left panels of Fig. \ref{dune_cp_3}.
But since the matter effect is more for antineutrinos, the precision of 
$\theta_{23}$ for IH and antineutrinos is poor (bottom right panel). 
\end{itemize} 
\item 
Similarly for $+90^\circ$-HO-IH although the octant is known the antineutrino run 
gives an enhanced $\chi^2$. This is due to matter effects in antineutrinos 
for IH which makes $P_{\bar{\mu} \bar{e}}$  higher than the corresponding 
neutrino probabilities (see Fig. \ref{oct_prob_deg}). This  
increases the appearance $\chi^2$ in presence of antineutrinos. 
For similar reasons the $\chi^2$ for 7+3 is slightly higher than 10+0 for $-90^\circ$-LO-IH

\end{itemize}

%
  
\begin{figure*}
        \begin{tabular}{lr}
               \hspace*{-0.4in}
                \includegraphics[width=0.45\textwidth]{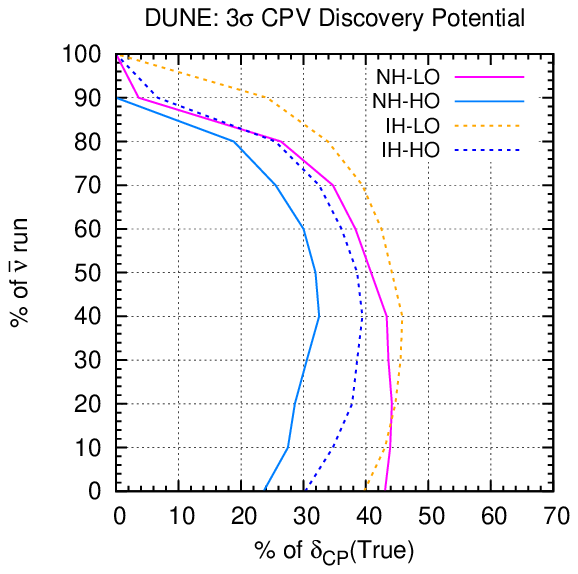}
                \hspace*{-1.0in}
                \includegraphics[width=0.45\textwidth]{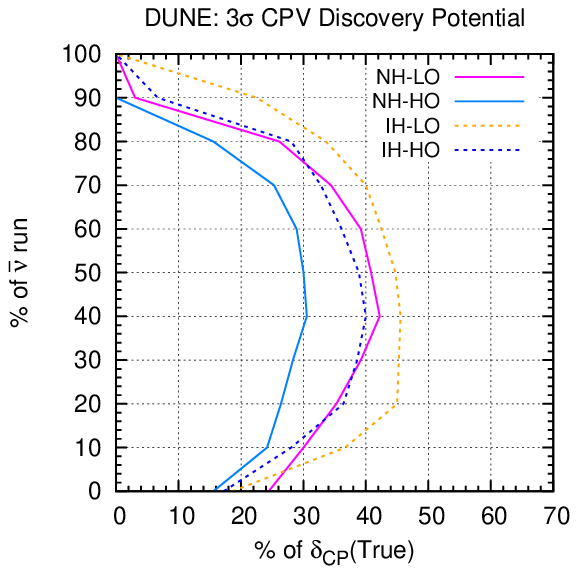}
                \hspace*{-1.0in}
                \includegraphics[width=0.45\textwidth]{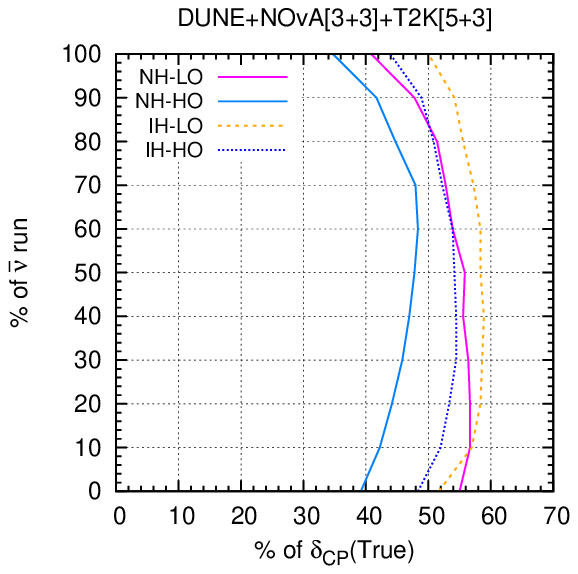}
        \end{tabular}
\vspace{-0.7cm}
\caption{ CP violation discovery $ \chi^{2}$ at 3$ \sigma $ C.L. in (\% of $ \delta_{CP}(True) $,  \% of antineutrino run) plane. Here, first and second column are for DUNE and third column is for  DUNE+NO$ \nu $A[3+3]+T2K[5+3]. Also y-axis represents the \% of antineutrino run out of total 10 years of [$ \nu + \overline{\nu} $] run in DUNE. In first (second and third) column we consider  hierarchy and octant are known (unknown).}
\label{dune_cpv_frac}
\end{figure*}
\begin{figure}
\hspace{-2.5cm}
        \begin{tabular}{lr}
               \hspace*{0.65in}
                \includegraphics[width=0.5\textwidth]{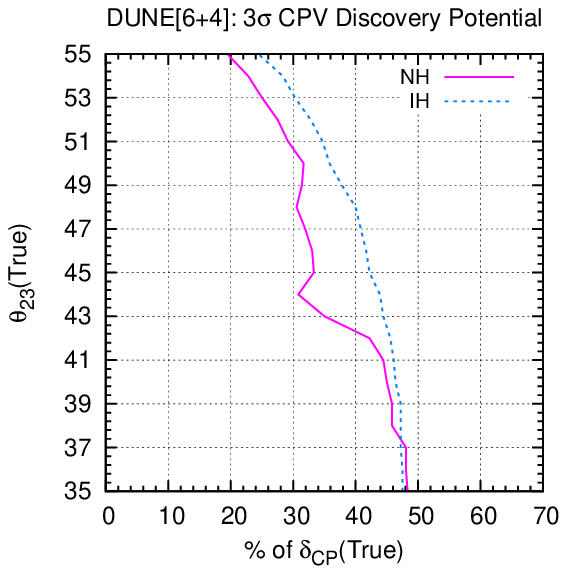}
        \end{tabular}
\vspace{-0.7cm}
\caption{CP violation discovery $ \chi^{2}$ at 3$ \sigma $ C.L. for DUNE[6+4] for all true $ \theta_{23} $ when hierarchy and octant are unknown. }
\label{dune_frac_th23}
\end{figure}

In the first and second  panels of figure \ref{dune_cpv_frac}, we plot the percentage of antineutrino run  vs percentage of $\dcp$ values  for which CP violated can be discovered at $3 \sigma$ C.L in DUNE for four cases encompassing both hierarchies and octants. The first (second) panel  represent when octant and hierarchy are known (unknown). From both plots it is seen that with dominant antineutrino or neutrino 
run a lesser CP fraction is reached. Overall 40\% antineutrino run seems to be optimum in all cases. 
Comparing these two plots it is seen that when octant is known then 
greater  percentage of CP fraction can be probed with less antineutrino 
component. The maximum CP coverage can be achieved for IH-HO and minimum for NH-HO. 

In the third panel of figure \ref{dune_cpv_frac} the same 
is plotted by combining \nova and T2K with DUNE. 
 From the figure we can see that the percentage of $\dcp$ that can be probed is enhanced in all cases. The curves are now much flatter implying  that even with pure neutrino 
or antineutrino runs considerable CP coverage can be obtained. 
This is due to the contribution from \nova and T2K. In figure \ref{dune_frac_th23} we show the dependence of 
percentage of $\dcp$ that can be probed as a function $\theta_{23}$.
This figure is drawn assuming 60\% neutrino and 40\% antineutrino run  
which is  the optimal configuration as seen in  Fig. \ref{dune_cpv_frac}. 
The coverage  of $\dcp$ for which CP violation can be discovered at $3 \sigma$ C.L is better for IH. For NH specially close to $45^\circ$  the coverage is less due to the poor precision of 
$\theta_{23}$ as discussed earlier.

%

\section{ Summary and Conclusions}
\label{sec5}

In this paper we perform  a detailed investigation of the 
octant and $\dcp$ sensitivity of the future generation 
superbeam experiment DUNE which has 
a baseline of 1300 km.
We analyze in detail the physics of the antineutrinos 
for the DUNE baseline and what kind of synergy can be offered by the 
addition of antineutrinos to pure neutrino runs. 
In the context 
of the long baseline experiments with source-detector 
distance $<1000$ km it  is well known that
the octant sensitivity comes mainly from the combination of  $P_{\mu e}$ 
and $P_{\mu \mu}$ channels. For $P_{\mu e}$ channel the $\chi^2$ is 
a rising function of $\theta_{23}$  and  consequently the 
minima in the wrong octant always  comes at $45^\circ$.  
On the other hand $P_{\mu \mu}$ being governed  by 
$\sin^2 2\theta_{23}$, the minima comes close to $\pi/2 - \theta_{23}$ 
with no octant sensitivity.     
When both channels are combined then the global minima comes closer 
to $\pi/2 - \theta_{23}$ where the appearance channel contributes
a large octant sensitive $\chi^2$.   
However the appearance channel is also 
affected by the occurrence of octant-$\dcp$ degeneracy  
which can lead to spurious solutions. 
The nature of this degeneracy 
is  the same for both  the hierarchies 
but has a complementary 
nature for neutrinos and antineutrinos i.e., the $\dcp$ and octant 
combination for which there is degeneracy in neutrinos is devoid of this
for antineutrinos. 
The upshot is that the combination of neutrino and  antineutrino
runs  helps to solve this degeneracy.  
On the other hand the statistics is more for neutrinos.
This leads to the question of what is the optimal combination 
of neutrino and antineutrino run for giving the maximum benefit 
for octant determination.  This issue has been addressed in this work 
in the context of the DUNE experiment. 
We also discuss to what extent  the  broad-band nature of the 
beam and enhanced matter effect influences the octant sensitivity 
and if any new features emerge as compared to the previous 
narrow-band off-axis experiments with baseline $<1000~ \mathrm{km}$.
We find that for the DUNE baseline addition of antineutrinos are helpful 
in general. This statement holds true  
even when there may be some degeneracy associated with the antineutrino channel
and one expects the pure neutrino run to give the best results.
This occurs because  of opposing tendencies of 
neutrino and antineutrino $\chi^2$s. 
We find that when $ \bar{\nu} $ is combined with $\nu$ then the overall
$ \chi^{2} $ minimum is still governed by the neutrinos because of 
higher statistics. At this point the antineutrino contribution to 
$\chi^2$ is higher  and hence adding these enhances octant sensitivity 
inspite of the associated octant degeneracy.  
One should also note that due to the broadband nature of the beam the degeneracy may be 
limited for few energy bins only. Note that, due to the broad-band nature of the beam, the octant sensitivity coming from  pure neutrino run is also
quite high at the true values where neutrino probabilities are themselves degenerate. 
%
The antineutrino contribution can be more for IH 
since due to enhanced matter effects the corresponding 
probabilities can be 
much higher than the neutrino probabilities. 
Thus even if the main octant sensitivity comes from the 
neutrinos, the broad-band nature compounded with the higher matter 
effect leads to some octant sensitivity coming from antineutrino channels
in case of IH.
In addition we find that a small octant sensitive contribution comes
from the disappearance channel when neutrino and antineutrino runs are 
combined although pure neutrino or pure antineutrino runs do not have this 
sensitivity. This happens because, 
due to matter effect the neutrino and antineutrino
probabilities are slightly different and hence the minima comes at 
slightly different position for each case. When combined, there is a tension
between these two which gives rise  to a small octant sensitive $\chi^2$  
contribution. 
Note that these features arising due to matter effects  
 have not been highlighted in the literature earlier.  

Taking two representative values of $\theta_{23}$ in the lower octant 
($39^\circ$) and higher octant ($51^\circ$) we study the behaviour of 
$\chi^2$ with $\dcp$ for both NH and IH.  
We find that for a 10 kt mass of the detector 
although for some $\dcp$ values $(3-4)\sigma$ sensitivity can be 
achieved with only neutrino run, overall adding antineutrinos is helpful. 
For a 7+3 year ($\nu + \bar{\nu}$) run, close to $4\sigma$ sensitivity can be achieved over all values of $\dcp$. It is found that 7+3 and 5+5 do not give 
significantly different results. 
We compute the $\chi^2$ as a function of true $\theta_{23}$ for maximum CP 
violation. From this study we find that with 7+3 years option octant 
degeneracy can be resolved at $3\sigma$  excepting the range $ 41.5^\circ < \theta_{23} < 49^\circ$.   Increasing the antineutrino component and making runtime
5+5 does not make any discernible difference to the results. 
Finally we also study the  octant sensitivity in the true($\theta_{23} - \dcp$) 
plane which checks the validity of the conclusions drawn earlier over the 
whole parameter range.  
We find that for 10 kt year mass the antineutrino run enhances the range of 
$\theta_{23}$ over which octant sensitivity can be achieved. 
Including antineutrino runs, octant sensitivity can be obtained  at $3\sigma$  
excepting the range $ 43^\circ < \theta_{23} < 49^\circ$ not only for 
maximal violation of $\dcp$ but over the whole range. 
In this case with only neutrino run  octant remains undetermined over 
a large  parameter space. 
We also present the 3$\sigma$ precision contours in the 
true $\theta_{23}$-test $\theta_{23}$ plane. These plots 
show that adding antineutrino runs also help in obtaining improved 
precision on $\theta_{23}$.   

We also present results on the CP  violation discovery potential of 
DUNE emphasizing the role played by the antineutrinos. 
The CP sensitivity of any long-baseline experiment is affected 
by the occurrence of the wrong-hierarchy-wrong octant-wrong $\dcp$ solutions. 
Since the antineutrinos help in removing these solutions, one of the 
main role of the antineutrinos in enhancing CP sensitivity is to 
remove these wrong solutions. We present results for 
cases where hierarchy and octant are unknown and known and compare 
the role of the antineutrinos in both situations. 
We find that when octant is  not known then in parameter spaces where 
octant degeneracy is manifest the antineutrino component increases 
CP sensitivity by removing wrong octant solutions. This is the case for instance for  LO, $\dcp \sim -90^\circ$ 
and HO, $\dcp \sim +90^\circ$ for both hierarchies. 
However even when the octant is known
addition of antineutrinos can improve the result because of the tension 
between the two $\chi^2$s which raises the overall $\chi^2$.  
The contribution from the antineutrino channel is higher for IH since due to 
matter effects the antineutrino probability is higher than the 
corresponding neutrino probability.  
At $\dcp = \pm 90^\circ$, a  greater than $3\sigma$ sensitivity is achieved in all cases. 
We have also explored how addition of  antineutrinos  
affects the fraction of $\dcp$ values 
for which CP sensitivity can be probed at $3\sigma$ level. 
We find that when octant is known, same sensitivity can be achieved
with a  lesser fraction of antineutrinos for both hierarchies. 
The maximum CP fraction is achieved for IH-LO. 
Overall the best result comes with 60\% neutrino 
and 40\% antineutrino runs for all the four cases.

In conclusion, we have explored the role of antineutrinos in 
enhancing  octant and CP sensitivity
for a 1300 km experiment with a broad-band beam as is planned by the DUNE 
collaboration. 
We emphasize on the importance of antineutrino run in resolving 
octant ambiguity and increasing CP sensitivity.  
Although  
for some specific parameters only neutrino run can give 
$3\sigma$ octant sensitivity for a 10 kt detector mass of DUNE,
 overall 
a balanced neutrino-antineutrino run gives better sensitivity. 
For  the case of $\dcp$ discovery also
in most of the parameter space antineutrinos  play 
an important role due to synergistic effects between neutrinos and
antineutrinos even under the assumption of octant to be known. 


\section{Acknowledgements}
 The authors would like to thank Sushant K. Raut for his help in GLoBES and also for many useful discussions regarding DUNE.

\bibliography{neutosc_new}{}
\end{document}